\documentclass[12pt,thmsa]{article}
\usepackage{amssymb}

\usepackage{sw20lart}

\input{tcilatex}
\begin{document}

\begin{center}
{\LARGE A prime factorization based on quantum dynamics on a spin ensemble
(I)} 
\[
\]
Xijia Miao*

\[
\]

Abstract
\end{center}

In this paper it has been described how to use the unitary dynamics of
quantum mechanics to solve the prime factorization problem on a spin
ensemble without any quantum entanglement. The ensemble quantum computation
for the prime factorization is based on the basic principle that both a
closed quantum system and its ensemble obey the same unitary dynamics of
quantum mechanics if there is not any decoherence effect in both the quantum
system and its ensemble. It uses the NMR multiple-quantum measurement
techniques to output the quantum computational results that are the inphase
multiple-quantum spectra of the spin ensemble. It has been shown that the
inphase NMR multiple-quantum spectral intensities used to search for the
period of the modular exponential function may reduce merely in a polynomial
form as the qubit number of the spin ensemble. The time evolution process of
the modular exponential operation on the quantum computer obeys the unitary
dynamics of quantum mechanics and hence the computational output is governed
by the quantum dynamics. This essential difference between the quantum
computer and the classical one could be the key point for the quantum
computation outperforming the classical one in the prime factorization on a
spin ensemble without any quantum entanglement. It has been shown that the
prime factorization based on the quantum dynamics on a spin ensemble is
locally efficient at least. This supports the conjecture that the quantum
dynamics could play an important role for the origin of power of quantum
computation and quantum entanglement could not be a unique resource to
achieve power of quantum computation in the prime factorization.\newline
---------------------------------------------------------

* Address: MSKCC, New York, NY 10017 and 9 Marney Street, Cambridge, MA
02141, USA. E\_mail: miaoxijia@yahoo.com 
\[
\]
\newline
{\large 1. Introduction}

The prime factorization is an important problem that has made it a rapid
development for the quantum computation and quantum information science. The
Shor$^{\prime }$s quantum algorithm [1, 2] proposed first in 1994 to
factorize efficiently a large composite integer can provide a possibility to
break down the current public key cryptography such as the RSA cryptosystem.
This fact has stimulated a great interesting in the quantum computation and
information science and has promoted a large advance in the quantum
computation [3, 4, 5, 6, 7]. This factorization algorithm could not come
along for a short time most because the present-day quantum systems have not
an enough long decoherence time to run the algorithm, although a preliminary
experiment verification for the Shor$^{\prime }s$ algorithm on an NMR
quantum computer was reported [8]. The factoring algorithm is based on a
pure-state quantum system. It has been suggested due to this powerful
quantum algorithm that the exponential speedup power of the quantum
computation over the classical counterpart could be attributed to quantum
entanglement of a quantum system [9]. One reason for it is that quantum
entanglement is a uniquely feature differing the quantum effect from the
classical effect and most powerful quantum algorithms nowadays involve in
the quantum entanglement [9]. Another is the well-known fact that quantum
entanglement plays a key important role in quantum communication [10, 11].
However, it has never been proved rigorously that quantum entanglement is
the sole origin of power of quantum computation, and a number of recent
works [12, 13] have showed that the power of quantum computation may not
originate from quantum entanglement, although no work shows so far that the
exponential speedup in the factoring algorithm may be independent of quantum
entanglement. Very recently, an improved factoring algorithm has been
proposed [14]. It has been shown that a quantum system consisting of an
auxiliary pure-state qubit and $\log _{2}N$ mixed qubits is still sufficient
to implement efficiently the prime factorization [14, 15]. But it also has
been argued [14, 16] that quantum entanglement could play an important role
in achievement of the exponential speedup in the algorithm since there still
exists quantum entanglement in such a system. The exponential speedup
achieved on such a system really does not provide any certain answer whether
or not quantum entanglement plays an important role in the exponential
speedup of quantum computation over the classical computation.

Fortunately, there are a lot of quantum ensembles in nature in which there
is not any quantum entanglement. These quantum ensembles include the
conventional NMR nuclear spin ensembles at room temperature [17], which are
also macroscopic quantum ensembles [18]. Quantum entanglement in a spin
ensemble may be controlled by temperature of the spin ensembles. It is easy
to keep any mixed state of a spin ensemble even with a larger number of
qubits in a nonentanglement state by setting the spin ensemble at a higher
temperature [17], but temperature of a spin ensemble should be as low as
possible in order to make the NMR signal-to-noise ratio high enough for any
NMR experiments. Such an ensemble without any quantum entanglement could be
a typical system to judge whether or not quantum entanglement is the origin
of power of quantum computation. Actually, if any quantum algorithm such as
the factoring algorithm could be implemented efficiently in such a quantum
ensemble without any quantum entanglement one could conclude certainly that
quantum entanglement is not the unique origin of power of quantum
computation. This is one of the reasons why the NMR spin ensembles are
chosen as the typical systems to study the origin of power of quantum
computation in the paper. Another reason is that the NMR spin ensembles
usually have a long relaxation time and are simple and easy to be controlled
and manipulated at will in experiments, and there are a large number of
well-developed experimental techniques in the NMR spectroscopy [19, 20]
which all can be adopted in NMR quantum computation.

The unitary dynamic method of quantum mechanics has been proposed to solve
efficiently the quantum search problem and the hard NP-problems [21, 22, 23,
24]. It has been shown that the unitary dynamics of quantum mechanics is the
base of ensemble quantum computation [22, 23, 24]. The basis principle
behind the ensemble quantum computation is that both a closed quantum system
and its ensemble obey the same unitary dynamics of quantum mechanics if
there is not any decoherence in both the quantum system and its ensemble.
This basis principle allows one to use quantum ensembles such as the spin
ensembles without any quantum entanglement to do real quantum computation.
Recently, the NMR multiple-quantum measurement techniques have been used to
output quantum computational results which are the inphase multiple-quantum
coherence spectra in a spin ensemble [24]. The NMR measurement for the
inphase multiple-quantum coherences need not an exponential resource when
the multiple-quantum coherences in the spin ensemble are created efficiently
by any quantum circuit. Therefore, both the unitary dynamics of quantum
mechanics and the multiple-quantum measurement techniques may form the base
for the scalable ensemble quantum computation [24]. In this paper both the
unitary dynamics of quantum mechanics and the multiple-quantum measurement
techniques have been exploited to solve the prime factorization problem on a
spin ensemble without any quantum entanglement. The purpose for it is to
study how the quantum dynamics plays an important role on the origin of
power of quantum computation in the prime factorization on a spin ensemble. 
\[
\]
\newline
{\large 2. Eigenvalues and eigenvectors of modular exponential }

{\large operation}

The integer factoring problem can be reduced to the order-finding problem,
while the latter is closely related to the unitary transformation of the
modular exponential operation in the quantum factoring algorithms based on
pure quantum states [5, 6]:

$\qquad \qquad U(y,r,N)|x\rangle =|xy\func{mod}N\rangle ,$ $%
x=0,1,...,N-1.\qquad \qquad \qquad (1)$ \newline
The transformation $U(y,r,N)$ is a unitary transformation only when the
numbers $y$ and $N$ are coprime to each other. The explicit form of the
unitary transformation is dependent only on the numbers $y$ and $N$. This
unitary transformation hides the period $r$ of the modular exponential
function $f(m)=f(y,m,N)=y^{m}\func{mod}N$ that is a periodic function: $%
f(m)=f(m+r)$. The period $r$ need to be determined in the order-finding
problem. The modular exponential unitary transformation (1) can be
efficiently implementable [1, 2, 5, 6]. Given a number $y$ prime to the
integer $N$ the unitary transformation $U(y,r,N)$ can be determined
explicitly from Eq.(1). Then the Hamiltonian corresponding to the unitary
operator $U(y,r,N)$ can be expressed in form

$\qquad \qquad \qquad H(y,r,N)=i\ln U(y,r,N).\qquad \qquad \qquad \qquad
\qquad \qquad \ \ (2)$ \newline
Since the order of the unitary operator $U(y,r,N)$ is $r$, i.e., $%
U(y,r,N)^{r}=E$ (the unity operator) there are $r$ different eigenvalues for
the unitary operator: $\Lambda _{k}=\exp (-i2\pi k/r),$ $k=0,1,...,r-1.$
Then the Hamiltonian $H(y,r,N)$ also have $r$ different eigenvalues: $%
\lambda _{k}=2\pi k/r,$ $k=0,1,...,r-1.$ According to the Cayley-Hamilton
theorem of linear algebra [25] the Hamiltonian $H(y,r,N)$ of Eq.(2)\ can be
expanded as

$\qquad \qquad \qquad H(y,r,N)=\stackrel{r-1}{\stackunder{k=0}{\sum }}\alpha
_{k}U(y,r,N)^{k}.\qquad \qquad \qquad \quad \qquad \qquad \ (3)$ \newline
Now suppose that the common eigenvectors of the unitary operator and its
Hamiltonian are denoted as $\{|\Psi _{k}\rangle \},$ then their
eigen-equations are respectively given by

$\qquad \qquad \qquad U(y,r,N)|\Psi _{k}\rangle =\Lambda _{k}|\Psi
_{k}\rangle $ \qquad \qquad \qquad \qquad \qquad $\quad \qquad (4a)$\newline
and

$\qquad \qquad \qquad H(y,r,N)|\Psi _{k}\rangle =\lambda _{k}|\Psi
_{k}\rangle .\qquad \qquad \qquad \qquad \quad \qquad \qquad (4b)$\newline
By the operator equation (3) and the eigenvector $|\Psi _{k}\rangle $ one
obtains from Eq.(4b)

$\qquad \qquad \qquad \lambda _{k}=\stackrel{r-1}{\stackunder{l=0}{\sum }}%
\alpha _{l}\exp (-i2\pi kl/r).\qquad \qquad \qquad \qquad \qquad \qquad \ \
(5a)$ \newline
Obviously, both the eigenvalues $\{\lambda _{k}=2\pi k/r\}$ and the
coefficients $\{\alpha _{k}\}$ form a pair of Fourier transform, and
according to the Fourier transform relation (5a) one can determine
explicitly the coefficients $\{\alpha _{k}\},$

$\qquad \qquad \qquad \alpha _{k}=\stackrel{r-1}{\stackunder{l=0}{\sum }}%
\frac{2\pi l}{r^{2}}\exp (i2\pi kl/r).\qquad \qquad \qquad \qquad \qquad
\quad \qquad (5b)$\newline
Then inserting Eq.(5b) into Eq.(3) the Hamiltonian $H(y,r,N)$ is written as

$\qquad \qquad H(y,r,N)=\stackrel{r-1}{\stackunder{k=0}{\sum }}\stackrel{r-1%
}{\stackunder{l=0}{\sum }}\frac{2\pi l}{r^{2}}\exp (i2\pi
kl/r)U(y,r,N)^{k}.\qquad \qquad \qquad (6)$\newline
The Hamiltonian $H(y,r,N)$ above is derived in detailed from its
corresponding unitary operator $U(y,r,N)$ of Eq.(1) partly due to that the
manipulation for a Hamiltonian is usually more convenient than for a unitary
operator in a complex spin ensemble [19, 20]. There is an important property
for the Hamiltonian $H(y,r,N)$ and unitary operator $U(y,r,N)$ of the
modular exponential operation (1) according to the definitions of the
unitary operator (1) and its Hamiltonian (2) that the unitary operator $%
U(y,r,N)^{m}$ can be written as

$\qquad U(y,r,N)^{m}=U(y^{m},r,N)=\exp [-imH(y,r,N)].\qquad \qquad \quad
\qquad (7)$\newline
The first equality in Eq.(7) shows that quantum circuit of the unitary
operation $U(y,r,N)^{m}$ can be efficiently constructed even when the
integer $m$ is a huge number, e.g., $m=r$, while the second equality shows
that the integer $m$ really acts as the discrete time variable in the
dynamical process of the modular exponential operation.

It is usually convenient to calculate time evolution of a spin ensemble
under a spin Hamiltonian if the eigenvectors and their eigenvalues of the
spin Hamiltonian are determined. The common eigenvectors of the unitary
operator $U(y,r,N)$ and its Hamiltonian $H(y,r,N)$ can be constructed
explicitly using the eigenequations (4a) and (4b). They should be a linear
combination of the conventional computational base $\{|k\rangle \}.$ Since
the period is $r$ for the modular exponential function $f(m)=f(y,m,N)=y^{m}%
\func{mod}N$, that is, there is the lowest integer $r$ such that $f(y,r,N)=1$%
, it follows from the unitary transformation (1) that for a given integer $x$
one has

$\quad \quad U(y,r,N)|xy^{m}\func{mod}N\rangle =|(xy^{(m+1)\func{mod}r})%
\func{mod}N\rangle ,\quad \quad \qquad \quad \quad \ \ \ (8)$ \newline
where $m=0,1,...,r-1.$ It is clear that the basis subset $\{|xy^{k}\func{mod}%
N\rangle ;$ $k=0,1,...,r-1\}$ form a closed state subset $S(x)$ under the
unitary transformation $U(y,r,N)$. For convenience, the dimension of the
state subset $S(x)$ is denoted as $r_{x}$ since it may depend on the integer 
$x.$ Obviously, for $x=1$ the dimension $r_{x}$ of the subset $S(x)$ equals $%
r$ exactly. The dimension $r_{x}$ is always smaller than or equal to the
period $r$ for any integer $x:$ $0\leq x<N-1$, that is, the period $r$ is
the maximum dimension in the subsets $S(x)$ for all possible $x,$ $0\leq
x<N-1$ also because according to the definition (1) the unitary operator $%
U(y,r,N)$ has the order $r$ and is independent of any $x$, $0\leq x<N-1$. In
particular, $r_{x}=1$ when $x=0$ and its subset $S(0)=\{|0\rangle \}$. It
follows from Eq.(8) that an arbitrary eigenstate $|\Psi _{s}(x)\rangle $ of
the unitary operator $U(y,r,N)$ of the subset $S(x)$ can be expressed as a
linear combination of the basis of the subset $S(x)$,

$|\Psi _{s}(x)\rangle =\stackrel{r_{x}-1}{\stackunder{k=0}{\sum }}%
c(r_{x},s,k)|xy^{k}\func{mod}N\rangle ,$ $s=0,1,..,r_{x}-1.\qquad \qquad \ \
\qquad (9)$\newline
Inserting the eigenstate (9) into the eigenequation (4a) one obtains

$\stackrel{r_{x}-1}{\stackunder{k=0}{\sum }}\exp (-i2\pi
s/r)c(r_{x},s,k)|xy^{k}\func{mod}N\rangle $

$\qquad \qquad \qquad -\stackrel{r_{x}-1}{\stackunder{k=0}{\sum }}%
c(r_{x},s,k)|xy^{k+1}\func{mod}N\rangle =0.\qquad \qquad \qquad \qquad \ \
(10)$ \newline
If dimension $r_{x}$ of the subset $S(x)$ is $r$ exactly then all $r$ states 
$\{|xy^{m}\func{mod}N\rangle ,$ $m=0,1,...,r-1\}$ are independent of each
other. Then the recursive relations for the coefficients $c(r,s,k)$ can be
set up by the eigenequation (10)

$\exp (-i2\pi s/r)c(r,s,k)=c(r,s,k-1),$ $k=1,2,...,r-1;$\newline
and

$\exp (-i2\pi s/r)c(r,s,0)=c(r,s,r-1).$\newline
Therefore, the coefficients $c(r,s,k)$ are determined by

$\qquad c(r,s,k)=\exp (i2\pi sk/r)c(r,s,0),$ $k=1,2,...,r-1.\qquad \qquad
\quad \ \ (11a)$ \newline
With the help of the orthonormal relations for the eigenstate $|\Psi
_{s}(x)\rangle $ and the basis $|xy^{k}\func{mod}N\rangle $: $\langle \Psi
_{s}(x)|\Psi _{s}(x)\rangle =1$ and $\langle xy^{k}\func{mod}N|xy^{k^{\prime
}}\func{mod}N\rangle =\delta _{kk^{\prime }}$ it is easy and straightforward
to find the coefficient $c(r,s,0)=$ $\frac{1}{\sqrt{r}}$ (here $c(r,s,0)$ is
taken as a real). Then the other coefficients are given explicitly by
Eq.(11a) once the coefficient $c(r,s,0)$ is known. Generally the modular
exponential function $f(x,y,m,N)=xy^{m}\func{mod}N$ ($x\neq 0$) may have the
same period $r$ as the function $f(y,m,N)=y^{m}\func{mod}N$, but besides the
period $r$ the function $f(x,y,m,N)$ may also have other periods $r_{x}$
different from $r$ for some given $x$ [14]. Then in the case of $r_{x}\leq r$
the eigenstate subset $\{|\Psi _{s}(x)\rangle \}$ of Eq.(9) of the unitary
operator $U(y,r,N)$ has both two period $r_{x}$ and $r$, and the period $%
r_{x}$ divides $r$ because there must be the relations: $|\Psi
_{0}(x)\rangle =|\Psi _{r}(x)\rangle =|\Psi _{lr_{x}}(x)\rangle $, $l$ is
some integer$.$ The eigenvalues of the unitary operator $U(y,r,N)$ belonging
to the eigenstates of Eq.(9) can be obtained using the eigenequation (4a),

$U(y,r,N)^{r_{x}}|\Psi _{s}(x)\rangle =\Lambda _{s}^{r_{x}}(x)|\Psi
_{s}(x)\rangle =|\Psi _{s}(x)\rangle ,$\newline
where the second equality is due to the period $r_{x}$, that is, $|xy^{k}%
\func{mod}N\rangle $ $=|xy^{k+r_{x}}\func{mod}N\rangle $. Therefore, the
eigenvalues are given by

$\Lambda _{s}(x)=\exp (-i2\pi s/r_{x}),$ $s=0,1,...,r_{x}-1.$\newline
Again using the eigenequation (4a) and the eigenvalues $\Lambda _{s}(x)$ one
can set up the recursive relations for the coefficients $c(r_{x},s,k)$ of
the eigenstate $|\Psi _{s}(x)\rangle $ similar to Eq.(11a) and hence the
coefficients $c(r_{x},s,k)$ are determined

$c(r_{x},s,k)=\frac{1}{\sqrt{r_{x}}}\exp (i2\pi sk/r_{x}),$ $%
k,s=0,1,...,r_{x}-1.\qquad \ \ \qquad \qquad (11b)$\newline
Then by using the coefficients of Eq.(11a) and Eq.(11b) one obtains from
Eq.(9) the common eigenstates of the unitary operator $U(y,r,N)$ and
Hamiltonian $H(y,r,N)$ [4, 6, 7]$,$

$|\Psi _{s}(x)\rangle =\frac{1}{\sqrt{r_{x}}}\stackrel{r_{x}-1}{\stackunder{%
k=0}{\sum }}\exp (i2\pi sk/r_{x})|xy^{k}\func{mod}N\rangle ,$ $%
s=0,1,..,r_{x}-1.\qquad (12a)$\newline
Obviously, the eigenstate set $\{|\Psi _{s}(x)\rangle \}$ and the basis set $%
\{|xy^{k}\func{mod}N\rangle \}$ form a pair of Fourier transforms, and the
inverse Fourier transform of Eq.(12a) generates the basis $|xy^{k}\func{mod}%
N\rangle $ as

$\qquad |xy^{k}\func{mod}N\rangle =\frac{1}{\sqrt{r_{x}}}\stackrel{r_{x}-1}{%
\stackunder{k=0}{\sum }}\exp (-i2\pi sk/r_{x})|\Psi _{s}(x)\rangle \qquad
\qquad \qquad \quad (12b)$ \newline
The Fourier transforms of Eq.(12a) and (12b) are helpful for calculating in
an analytical form the time evolution of a spin ensemble under the unitary
operation $U(y,r,N)$. This can be seen in next sections.

Since every integer $k$ in $0\leq k<N$ always can be expressed as $k=xy^{m}%
\func{mod}N$ by choosing suitably the integers $x$ and $m,$ where the
integer $y$ is coprime to and smaller than the integer $N$, then the
conventional computational basis can be expressed either as $\{|k\rangle ,$ $%
k=0,1,2,...,N-1\}$ or as $\{|xy^{m}\func{mod}N\rangle ,$ $%
m=0,1,2,...,r_{x}-1 $; $x=0,1,...,N-1\}.$ One can classify the conventional
computational basis $\{|k\rangle \}$ or $\{|xy^{m}\func{mod}N\rangle \}$
according to the transformation property of the unitary operator $U(y,r,N)$.
For example, for a given integer $x$ one can generate a basis subset $%
S(x)=\{|xy^{m}\func{mod}N\rangle ,$ $m=0,1,...,r_{x}-1\}.$ The whole Hilbert
state space with dimension $N$ then is divided into $t$ independent and
orthogonal basis subsets $S(x)$ with different integers $x:x_{0}\leq
x_{1}\leq ...\leq x_{t-1}.$ Clearly, $N=r_{x_{0}}+r_{x_{1}}+...+r_{x_{t-1}}.$
Therefore, the conventional computational basis set also can be expressed in
the simpler form $\{|x_{l}y^{k}\func{mod}N\rangle ,$ $k=0,1,...,r_{x_{l}}-1;$
$l=0,1,...,t-1\}$ according to the transformation property of the unitary
operator $U(y,r,N)$ (1), which is also equivalent to the conventional
computational basis set $\{|k\rangle \}.$ In particular, for $x_{0}=0$ the
subset $S(x_{0})=\{|0\rangle \}$ with $r_{x_{0}}=1$ and for $x_{1}=1$ the
subset $S(x_{1})$ is an $r-$dimensional subset. Actually, besides the subset 
$S(x_{1})$ there may be also other $r-$dimensional subsets $S(x)$ with $x>1.$
Suppose that there are $d$ independent $r-$dimensional subsets including $%
x_{1}=1$ and $x>1$ in the $N-$dimensional Hilbert space$.$ There are $rd$
computational basis that belong to the $r-$dimensional subsets among the $N$
computational basis, and the rest $N-rd$ basis are of those subsets with
dimensions $r_{x}$ smaller than $r$. How many the computational basis belong
to the $r-$dimensional subsets in all $N$ computational basis? This can be
answered by the theorem (Parker and Plenio [14]): Given two prime numbers $p$
and $q$, $N=pq$, $r$ is defined as the period of the modular exponential
function $f(m)=f(y,m,N)=$ $y^{m}\func{mod}N$ for an arbitrary integer $y$,
then there are at least $(p-1)(q-1)$ positive integers $x$ less than and
coprime to the integer $N$ such that the modular exponential function $%
g(m)=f(x,y,m,N)=xy^{m}\func{mod}N$ has the minimum period equal to $r$ for $%
0\leq y\leq N-1.$

Actually, the unitary transformation (1) shows that the conventional
computational basis set $\{|k\rangle \}$ is also equivalent to the basis set 
$\{|xy^{k}\func{mod}N\rangle ,$ $x=0,1,...,N-1\}$ for any given integer $k$.
Then according to the theorem [14] number of the computational basis that
satisfy $xy^{k+r_{x}}=xy^{k}\func{mod}N$ with $r_{x}<r$ is $(p+q-1)$ at most
in the whole basis set $\{|xy^{k}\func{mod}N\rangle ,$ $x=0,1,...,N-1\}$.
This also means that number of the computational basis that belong to those
basis sets $S(x)$ with dimensions $r_{x}<r$ is at most $(p+q-1)$ in the
complete basis set $\{|x_{l}y^{k}\func{mod}N\rangle ,$ $%
k=0,1,...,r_{x_{l}}-1;$ $l=0,1,...,t-1\}.$ Therefore, the number $N-rd$ of
the computational basis of those subsets with dimensions $r_{x}$ smaller
than $r$ is at most $(p+q-1)$, and the number $rd$ of the computational
basis of the $r-$dimensional subsets in the $N-$dimensional ($N=pq$) Hilbert
space is at least $pq-(p+q-1).$

Finally, it is also important to know the orthonormal relations for the
eigenstates $|\Psi _{s}(x_{k})\rangle $ and the conventional computational
basis $|x_{l}y^{k}\func{mod}N\rangle $ for conveniently calculating the time
evolution of a spin ensemble under the unitary operation $U(y,r,N)$,

$\qquad \qquad \langle \Psi _{k}(x_{l})|\Psi _{k^{\prime }}(x_{l^{\prime
}})\rangle =\delta _{kk^{\prime }}\delta _{ll^{\prime }},\qquad \qquad
\qquad \qquad \qquad \quad \qquad \ \ (13a)$

$\qquad \qquad \langle x_{l}y^{k}\func{mod}N|x_{l^{\prime }}y^{k^{\prime }}%
\func{mod}N\rangle =\delta _{kk^{\prime }}\delta _{ll^{\prime }}.\qquad
\qquad \qquad \qquad \quad \ \ (13b)$ \newline
where $k,k^{\prime }=0,1,...,r_{x_{l}}-1$ and $l,l^{\prime }=0,1,...,t-1.$

\[
\]
\newline
{\large 3. Time evolution process of modular exponential operation }

Generally the prime number $N$ is not equal to some power of two. Suppose
that the prime number $N=pq$ satisfies $2^{n-1}\leq N<2^{n}.$ For
simplifying calculation of the time evolution of a spin ensemble during the
modular exponential operation the NMR quantum computer could be chosen
conveniently as a heteronuclear spin ensemble $%
I_{1}I_{2}...I_{n_{i}}S_{1}S_{2}...S_{n}$ (denoted briefly as $%
I_{n_{i}}S_{n} $) that consists of $n_{i}$ spin-1/2 $I$ nuclei and $n$
spin-1/2 $S$ nuclei and particularly $n_{i}=1,2,...$. This is just like the
pure-state Shor$^{\prime }s$ factoring algorithm using two memories [1, 2].
The conditional modular exponential operation $U_{I_{k}S_{n}}(y,r,N)$
applying to the spin system $I_{n_{i}}S_{n}$ is built up with the unitary
transformation of Eq.(1),\newline
$U_{I_{k}S_{n}}(y,r,N)|a\rangle |x\rangle =\{ 
\begin{array}{l}
|a\rangle (U(y,r,N)|x\rangle ),\text{ }x=0,1,...,N-1;\text{ }a=1 \\ 
|a\rangle |x\rangle ,\text{ }x=0,1,...,N-1;\text{ }a=0 \\ 
|a\rangle |x\rangle ,\text{ }x=N,N-1,...,2^{n}-1;\text{ }a=0,1
\end{array}
\quad \ (14)$ \newline
where the quantum states $|a\rangle $ and $|x\rangle $ belong to the $k$th
spin $I$ and all $n$ spins $S_{n}$ of the spin system $I_{n_{i}}S_{n}$,
respectively. The modular exponential operation $U(y,r,N)$ is applied only
to those quantum state $|x\rangle $ of the spin subsystem $S_{n}$ with $x<N$
only if the $k$th $I-$spin quantum state $|a\rangle =|1\rangle .$ It can
turn out that the Hamiltonian corresponding to the unitary operator $%
U_{I_{k}S_{n}}(y,r,N)$ (14) may be expressed as

$H_{I_{k}S_{n}}(y,r,N)=E_{1}\bigotimes ...\bigotimes E_{k-1}$

$\qquad \bigotimes (\frac{1}{2}E_{k}-I_{kz})\bigotimes E_{k+1}\bigotimes
...E_{n_{1}}\bigotimes H_{S_{n}}(y,r,N)\qquad \qquad \qquad \qquad (15)$%
\newline
where $I_{kz}|0\rangle =\frac{1}{2}|0\rangle $ $(\hslash =1)$ and $%
I_{kz}|1\rangle =-\frac{1}{2}|1\rangle $, and the Hamiltonian $%
H_{S_{n}}(y,r,N)$ applied only to the subsystem $S_{n}$ is defined as

$\qquad \qquad H_{S_{n}}(y,r,N)=H(y,r,N)\bigoplus Z_{L-N}\qquad \qquad
\qquad \qquad \qquad \quad (16)$\newline
where $Z_{L-N}$ is the $(L-N)\times (L-N)-\dim $ensional zero operator$.$ It
is easy to prove according to Eq.(7) that the conditional modular
exponential operation $U_{I_{k}S_{n}}(y,r,N)^{m}$ can be expressed as

$\qquad U_{I_{k}S_{n}}(y,r,N)^{m}=U_{I_{k}S_{n}}(y^{m},r,N)=\exp
[-imH_{I_{k}S_{n}}(y,r,N)].\qquad (17)$\newline
The unitary operator $U_{I_{k}S_{n}}(y,r,N)^{m}$ can be constructed
efficiently for any integer $m$ because the unitary operator $%
U_{I_{k}S_{n}}(y^{m},r,N)$ can be constructed efficiently. It is known from
Eq.(17) that the power $m$ in the unitary operator $%
U_{I_{k}S_{n}}(y,r,N)^{m} $ really acts as the discrete time variable in the
dynamical process of the conditional modular exponential operation. Note
that any pair of the Hamiltonians $H_{I_{k}S_{n}}(y,r,N)$ (15) with
different I-spin operators $I_{kz}$ $(k=1,2,...,n_{i})$ commute each other.
A more general conditional modular exponential operation can be constructed
by

$\qquad \qquad U_{I_{n_{i}}S_{n}}(y,r,N)=\stackrel{n_{1}}{\stackunder{k=1}{%
\prod }}U_{I_{k}S_{n}}(y,r,N),\qquad \qquad \qquad \qquad \qquad \quad (18)$%
\newline
and it is easy to prove that the relation (17) also is met for the general
conditional modular exponential operation (18). The Hamiltonian of the
unitary operator $U_{I_{n_{i}}S_{n}}(y,r,N)$ therefore is written as

$\qquad \qquad H_{I_{n_{i}}S_{n}}(y,r,N)=\stackrel{n_{1}}{\stackunder{k=1}{%
\sum }}H_{I_{k}S_{n}}(y,r,N).\qquad \qquad \qquad \qquad \qquad \ \ (19)$

Obviously, the conditional unitary operation $U_{I_{k}S_{n}}(y,r,N)$ is
independent of any quantum state $|a\rangle |x\rangle $ (as the initial
input state) of the spin system $I_{n_{i}}S_{n}$. This suggests that the
unitary operator can be applied not only to any pure quantum states of the
spin system $I_{n_{i}}S_{n}$ but also directly to any mixed states of the
spin ensemble $I_{n_{i}}S_{n}$ of the spin system [22, 23]. This is just the
essence of the basic principle that both a closed quantum system and its
ensemble obey the same unitary dynamics of quantum mechanics if there is not
any decoherence effect in both the quantum system and its ensemble [22, 23,
24]. This principle forms the base of the current factoring algorithm and
the real implementation of the algorithm on a spin ensemble. The unitary
operator $U_{I_{k}S_{n}}(y,r,N)$ hides the period $r$ to be determined. In
order to find the period it first needs to transfer the information of the
period $r$ of the unitary operator into quantum states of a quantum system
or the density operator of its quantum ensemble because both the quantum
states and density operators can be measured conveniently in practice.
According to the basic principle the initial input state of the unitary
operator $U_{I_{k}S_{n}}(y,r,N)$ can take either any pure quantum state of a
closed quantum system or any mixed state, i.e., density operator of its
ensemble, but in an NMR spin ensemble it is most convenient to take the
initial density operator, i.e., the input state of the current factoring
algorithm, as the thermal equilibrium state of the spin ensemble. For the
spin ensemble $I_{n_{i}}S_{n}$ in a high magnetic field the thermal
equilibrium state can be written as, in high temperature approximation

$\qquad \qquad \qquad \rho _{eq}=\alpha E+\stackrel{n_{1}}{\stackunder{k=1}{%
\sum }}\varepsilon _{ik}I_{kz}+\stackrel{n}{\stackunder{k=1}{\sum }}%
\varepsilon _{sk}S_{kz}\qquad \qquad \qquad \qquad \quad \ \ (20)$ \newline
where $E$ is the unity operator and the operators $I_{kz}$ and $S_{kz}$ are
the longitudinal magnetization operators of the $k$th spins $I$ and $S,$
respectively. Then a nonselective $90_{y}^{\circ }$ excitation pulse $%
R_{i}(90_{y}^{\circ })=\exp (-i\pi I_{y}/2)$ applied to all the spins $I$
and a $90_{\varphi }^{\circ }$ nonselective pulse $R_{s}(90_{\pm y}^{\circ
})=\exp [\mp i(\pi /2)S_{y}]$ with two-step phase cycling $\varphi =+y,-y$
applied to all the spins $S$ convert the thermal equilibrium state (20) into
the single-quantum density operator,

$\qquad \qquad \rho (0)=(\stackrel{n_{1}}{\stackunder{k=1}{\sum }}%
\varepsilon _{ik}I_{kx})\bigotimes E_{1}^{s}\bigotimes E_{2}^{s}\bigotimes
...\bigotimes E_{n}^{s}\qquad \qquad \qquad \quad \ \ (21)$ \newline
where the unity operator term $\alpha E$ is neglected without losing
generality and $E_{k}^{s}$ is the $4-$dimensional unity operator of the $k$%
th spin $S$. The two-step phase cycling $\varphi =+y,-y$ [19, 20] cancels
the contribution of the thermal equilibrium magnetization $\rho _{seq}=%
\stackrel{n}{\stackunder{k=1}{\sum }}\varepsilon _{sk}S_{kz}$ of the
subensemble $S_{n}$ of the spin ensemble $I_{n_{i}}S_{n}$ to the output NMR
signal, leaving only the thermal equilibrium magnetization $\rho _{ieq}=%
\stackrel{n_{1}}{\stackunder{k=1}{\sum }}\varepsilon _{ik}I_{kz}$ of the
subensemble $I_{n_{i}}$ having a net contribution to the output NMR signal.
Now the information of the period $r$ in the unitary operator $%
U_{I_{n_{i}}S_{n}}(y,r,N)$ need to be loaded on the density operator of the
spin ensemble $I_{n_{i}}S_{n}.$ This can be achieved by applying the unitary
operator on the initial density operator $\rho (0).$ An analytical
calculation is important for the time evolution of the spin ensemble with
the initial density operator $\rho (0)$ under the conditional unitary
operation $U_{I_{n_{i}}S_{n}}(y,r,N)$. It can be performed conveniently by
first expressing the initial density operator (21) in terms of the common
eigenvectors of the unitary operator $U(y,r,N)$ (1) and its Hamiltonian $%
H(y,r,N)$ (2). By using the conventional computational basis set $%
\{|x_{l}y^{k}\func{mod}N\rangle ,$ $k=0,1,...,r_{x_{l}}-1;$ $l=0,1,...,t-1\}$
the initial density operator (21) is rewritten as

$\rho (0)=(\stackrel{n_{1}}{\stackunder{k=1}{\sum }}\varepsilon
_{ik}I_{kx})\bigotimes \stackrel{t-1}{\stackunder{l=0}{\sum }}\stackrel{%
r_{x_{l}}-1}{\stackunder{k=0}{\sum }}|x_{l}y^{k}\func{mod}N\rangle \langle
x_{l}y^{k}\func{mod}N|$

$\qquad +(\stackrel{n_{1}}{\stackunder{k=1}{\sum }}\varepsilon
_{ik}I_{kx})\bigotimes \stackrel{L-1}{\stackunder{k_{s}=N}{\sum }}%
|k_{s}\rangle \langle k_{s}|$,$\quad (L=2^{n},$ $\frac{1}{2}L<N<L).\qquad
\qquad \ (22)$\newline
With the help of the inverse Fourier transform of Eq.(12b) and the
orthonormal relation (13b) the density operator $\rho (0)$ is further
expressed as

$\rho (0)=(\stackrel{n_{1}}{\stackunder{k=1}{\sum }}\varepsilon
_{ik}I_{kx})\bigotimes \stackunder{k=0}{\stackrel{t-1}{\sum }}\stackrel{%
r_{x_{k}}-1}{\stackunder{s=0}{\sum }}|\Psi _{s}(x_{k})\rangle \langle \Psi
_{s}(x_{k})|$

$\qquad \qquad +(\stackrel{n_{1}}{\stackunder{k=1}{\sum }}\varepsilon
_{ik}I_{kx})\bigotimes \stackrel{L-1}{\stackunder{k_{s}=N}{\sum }}%
|k_{s}\rangle \langle k_{s}|.$\qquad $\qquad \qquad \qquad \qquad \qquad
\quad \ (23)$ \newline
On the other hand, with the help of the definition (14) of the conditional
unitary operation $U_{I_{k}S_{n}}(y,r,N)$ and its the Hamiltonian (15) as
well as the eigenequation (4b) of the Hamiltonian $H(y,r,N)$ it is now easy
to calculate the time evolution of the spin ensemble when applying the
conditional unitary operation $U_{I_{n_{i}}S_{n}}(y,r,N)^{m}$ on the density
operator $\rho (0)$ of Eq.(23),

$\rho (m)=U_{I_{n_{i}}S_{n}}(y,r,N)^{m}\rho
(0)U_{I_{n_{i}}S_{n}}^{+}(y,r,N)^{m}$

$=-\varepsilon _{i}I_{y}\bigotimes \stackunder{l=0}{\stackrel{t-1}{\sum }}%
\stackrel{r_{x_{l}}-1}{\stackunder{s=0}{\sum }}\sin [2\pi ms/r_{x_{l}}]|\Psi
_{s}(x_{l})\rangle \langle \Psi _{s}(x_{l})|$

$+\varepsilon _{i}I_{x}\bigotimes \stackunder{l=0}{\stackrel{t-1}{\sum }}%
\stackrel{r_{x_{l}}-1}{\stackunder{s=0}{\sum }}\cos [2\pi ms/r_{x_{l}}]|\Psi
_{s}(x_{l})\rangle \langle \Psi _{s}(x_{l})|$

$+\varepsilon _{i}I_{x}\bigotimes \stackrel{L-1}{\stackunder{k_{s}=N}{\sum }}%
|k_{s}\rangle \langle k_{s}|,\qquad \qquad \qquad \qquad \qquad \qquad
\qquad \qquad \qquad \qquad \ (24)$ \newline
where all $n_{i}$ spins $I$ have the same spin polarization factor $%
\varepsilon _{i},$ that is, $\varepsilon _{i}I_{\mu }=\stackrel{n_{1}}{%
\stackunder{k=1}{\sum }}\varepsilon _{ik}I_{k\mu }$ $(\mu =x,y,z).$ One can
obtain the antisymmetric $y-$component of the density operator (24),

$\quad \rho _{y}(m)=-\varepsilon _{i}I_{y}\bigotimes \stackunder{l=0}{%
\stackrel{t-1}{\sum }}\stackrel{r_{x_{l}}-1}{\stackunder{s=0}{\sum }}\sin
[2\pi ms/r_{x_{l}}]|\Psi _{s}(x_{l})\rangle \langle \Psi _{s}(x_{l})|,\qquad
\qquad \ (25)$ \newline
by doing another experiment: $\rho
(-m)=U_{I_{n_{i}}S_{n}}^{+}(y,r,N)^{m}\rho (0)U_{I_{n_{i}}S_{n}}(y,r,N)^{m}$
and then coadding coherently the final output NMR signals of the two density
operators $\rho (m)$ and $[-\rho (-m)]$, that is, $\rho _{y}(m)=\frac{1}{2}%
(\rho (m)-\rho (-m)).$ The inverse unitary operation $%
U_{I_{n_{i}}S_{n}}^{+}(y,r,N)^{m}$ can be also implemented efficiently just
like the unitary operation $U_{I_{n_{i}}S_{n}}(y,r,N)^{m}$, as can be seen
in next section. The density operator $\rho _{y}(m)$ of Eq.(25) is
antisymmetric for any integer $m:$ $\rho _{y}(kr/2-m)=-\rho _{y}(kr/2+m),$ $%
k=0,1,2,...,$ because the period $r$ can be divided by any dimensions $%
r_{x_{l}}$. By inserting the eigenstates $|\Psi _{s}(x_{l})\rangle $ of
Eq.(12a) into Eqs.(24) and (25)\ one can express the density operators $\rho
(m)$ and $\rho _{y}(m)$ in terms of the conventional computational basis,

$\rho (m)=-\varepsilon _{i}I_{y}\bigotimes \stackunder{l=0}{\stackrel{t-1}{%
\sum }}\stackrel{r_{x_{l}}-1}{\stackunder{k=0}{\sum }}(\frac{1}{2i}%
)\{-|x_{l}y^{k}\func{mod}N\rangle \langle x_{l}y^{k+m}\func{mod}N|$

$\qquad +|x_{l}y^{k+m}\func{mod}N\rangle \langle x_{l}y^{k}\func{mod}%
N|\}\qquad $

$\qquad +\varepsilon _{i}I_{x}\bigotimes \stackunder{l=0}{\stackrel{t-1}{%
\sum }}\stackrel{r_{x_{l}}-1}{\stackunder{k=0}{\sum }}(\frac{1}{2}%
)\{|x_{l}y^{k}\func{mod}N\rangle \langle x_{l}y^{k+m}\func{mod}N|$

$\qquad +|x_{l}y^{k+m}\func{mod}N\rangle \langle x_{l}y^{k}\func{mod}%
N|\}+\varepsilon _{i}I_{x}\bigotimes \stackrel{L-1}{\stackunder{k_{s}=N}{%
\sum }}|k_{s}\rangle \langle k_{s}|\qquad \ \qquad \ \ \ (26)$ \newline
and

$\rho _{y}(m)=-\varepsilon _{i}I_{y}\bigotimes \stackunder{l=0}{\stackrel{t-1%
}{\sum }}\stackrel{r_{x_{l}}-1}{\stackunder{k=0}{\sum }}(\frac{1}{2i}%
)\{-|x_{l}y^{k}\func{mod}N\rangle \langle x_{l}y^{k+m}\func{mod}N|$

$\qquad \qquad +|x_{l}y^{k+m}\func{mod}N\rangle \langle x_{l}y^{k}\func{mod}%
N|\}.\qquad \qquad \qquad \qquad \qquad \qquad (27)$

Note that the conventional computational basis $\{|x_{l}y^{k}\func{mod}%
N\rangle ,l=0,1,...,$ $t-1;k=0,1,...,r_{x_{l}}-1\}$ of the subensemble $%
S_{n} $ are orthogonal to each other, as shown in Eq.(13b). Then the
operator $|x_{l}y^{k}\func{mod}N\rangle \langle x_{l}y^{k+m}\func{mod}N|$ is
a diagonal operator only when the equality $x_{l}y^{k}\func{mod}%
N=x_{l}y^{k+m}\func{mod}N$ holds. As shown in the previous section, the
equality holds only when $m=k^{\prime }r_{x_{l}}$ $(k^{\prime }=0,1,...,).$
Therefore, the operator $|x_{l}y^{k}\func{mod}N\rangle \langle x_{l}y^{k+m}%
\func{mod}N|$ is an off-diagonal operator when $m\neq k^{\prime }r_{x_{l}}$.
It is known that the diagonal elements of a density operator are the
noncoherence components which are known as the conventional longitudinal
magnetization and spin order components in NMR spectroscopy, while the
off-diagonal elements represent the coherent components of the density
operator, and it is also well known in NMR spectroscopy that the coherent
components of a density operator are the conventional multiple-quantum
coherences including single-quantum coherence [19, 20]. It is shown below
that the density operator $\rho _{y}(m)$ of Eq.(27) is a pure
multiple-quantum coherence operator of the subensemble $S_{n}$. Here assume
that the period $r$ is an even integer. First, the density operator $\rho
_{y}(m)$ is clearly a pure multiple-quantum coherences when $m\neq k^{\prime
}r_{x_{l}}.$ Next, it need to be shown that the integers $m=k^{\prime
}r_{x_{l}}$ are the zero points of the density operator $\rho _{y}(m).$ The
zero points of the density operator $\rho _{y}(m)$ are defined as those
integers $m$ satisfying $\rho _{y}(m)=0.$ Obviously, $m=k^{\prime }r$ are
the zero points of the density operator $\rho _{y}(m)$ because the period $r$
can be divided by any dimensions $r_{x_{l}}$ and hence $m=k^{\prime
}r=k_{x_{l}}r_{x_{l}}$ ($k_{x_{l}}$ is an integer) which lead to the
identity:

$\stackrel{r_{x_{l}}-1}{\stackunder{k=0}{\sum }}[|x_{l}y^{k+m}\func{mod}%
N\rangle \langle x_{l}y^{k}\func{mod}N|$

$\ \ \qquad -|x_{l}y^{k}\func{mod}N\rangle \langle x_{l}y^{k+m}\func{mod}%
N|]=0,$ $(l=0,1,...,t-1),\qquad \quad \ \ (28)$\newline
and thus, the density operator $\rho _{y}(m)=0$. Moveover, it is easy to
prove that every integer $m=(2k^{\prime }+1)r/2$ satisfies the formula (28)
and hence $m=(2k^{\prime }+1)r/2$ are also the zero points of the density
operator $\rho _{y}(m).$ Therefore, all possible zero points of the density
operator $\rho _{y}(m)$ are given by $m=k^{\prime }r/2$ $(k^{\prime
}=0,1,2,...)$. Then for any integer $m$ the density operator $\rho _{y}(m)$
is either an off-diagonal operator or equal to zero. This indicates that the
density operator $\rho _{y}(m)$ is a pure multiple-quantum coherence
operator of the subensemble $S_{n}$ and does not contain any longitudinal
magnetization and spin order ($LOMSO$) operators of the spin subensemble.

The density operator $\rho (m)$ of Eq.(26) is more complicated and consists
of both the $LOMSO$ operators and multiple-quantum coherence operators of
the subensemble $S_{n}$. The first term on the right-hand side of Eq.(26) is
really the pure multiple-quantum coherence operator $\rho _{y}(m)$, while
the last term really consists of pure $LOMSO$ operators. These $LOMSO$
operators are really invariants under the unitary transformation in Eq.(24)
with the unitary operator $U_{I_{n_{i}}S_{n}}(y,r,N)^{m}.$ The second term
contains the invariant diagonal element $\rho _{00}=|0\rangle \langle 0|$
which keeps unchanged under the unitary transformation. When the integers $%
m=k^{\prime }r$ the density operator of Eq.(26) equals the initial density
operator: $\rho (m)=\varepsilon _{i}I_{x}=\rho (0),$ indicating that at the
points $m=k^{\prime }r$ the initial density operator $\rho (0)$ is not
transferred into any multiple-quantum coherences of the subensemble $S_{n}$
under the unitary transformation$.$ Actually, if $m=k^{\prime }r$ the
unitary operator $U_{I_{n_{i}}S_{n}}(y,r,N)^{m}$ is the unity operator, that
is, $U_{I_{n_{i}}S_{n}}(y,r,N)^{m}=E$, and any initial density operator
keeps unchanged by the unity operation. One sees that at $m=k^{\prime }r$
there is not any multiple-quantum coherence of the subensemble $S_{n}$ in
the density operator $\rho (m)$ and also in the density operator $\rho
_{y}(m)$. Therefore, these integers $m=k^{\prime }r$ $(k^{\prime }=0,1,...,)$
are really the zero points of the density operators $\rho (m)$ and $\rho
_{y}(m)$ in the sense that there is not any multiple-quantum coherence of
the subensemble $S_{n}$ in these density operators$,$ although $\rho
(m)=\rho (0)\neq 0$ at $m=kr$. When $m=kr_{x_{l}}$ the whole conventional
computational basis subset $S(x_{l})=\{|x_{l}y^{k}\func{mod}N\rangle ,$ $%
k=0,1,...,r_{x_{l}}-1\}$ of the subensemble $S_{n}$ keeps unchanged under
the unitary operation $U_{I_{n_{i}}S_{n}}(y,r,N)^{m}$, showing that the $%
r_{x_{l}}$ operator terms $(\stackrel{n_{1}}{\stackunder{k=1}{\sum }}%
\varepsilon _{ik}I_{kx})\bigotimes \stackrel{r_{x_{l}}-1}{\stackunder{k=0}{%
\sum }}|x_{l}y^{k}\func{mod}N\rangle \langle x_{l}y^{k}\func{mod}N|$ of the
initial density operator $\rho (0)$ of Eq.(22) can not be transferred into
the multiple-quantum coherences. Based on these facts one can calculate the
conversion efficiency of the initial density operator $\rho (0)$ into the
multiple-quantum coherences under the unitary transformation.

It follows from Eq.(21) that the initial density operator $\rho (0)$
consists of $L$ operator terms $\rho _{k_{s}}=(\stackrel{n_{1}}{\stackunder{%
k=1}{\sum }}\varepsilon _{ik}I_{kx})\bigotimes |k_{s}\rangle \langle k_{s}|,$
$k_{s}=0,1,...,L-1.$ That the density operator $\rho (0)$ can not be
transferred completely into the multiple-quantum coherences of the
subensemble $S_{n}$ is because the $L-N$ operator terms $\stackrel{L-1}{%
\stackunder{k_{s}=N}{\sum }}\rho _{k_{s}}$ ($N\geq L/2$) and $\rho _{00}=(%
\stackrel{n_{1}}{\stackunder{k=1}{\sum }}\varepsilon _{ik}I_{kx})|0\rangle
\langle 0|$ of the density operator $\rho (0)$ keeps unchanged under the
conditional unitary operation $U_{I_{n_{i}}S_{n}}(y,r,N)^{m}$ with any
integer $m$. But the total contribution of these invariant operator terms to
the density operator $\rho (0)$ is clearly less than 50\%. Therefore, when $%
m\neq kr_{x_{l}}$ ($l=0,1,...,t-1$) the initial density operator $\rho (0)$
is transferred into the multiple-quantum coherences in a high efficiency of
50\% at least, and this efficiency is independent of the qubit number $n$ of
the spin ensemble ($I_{n_{i}}S_{n}$). For the case $m=kr_{x_{l}}$ $($each $%
r_{x_{l}}<r)$ but $m\neq k^{\prime }r$ the $r_{x_{l}}$ operator terms $(%
\stackrel{n_{1}}{\stackunder{k=1}{\sum }}\varepsilon _{ik}I_{kx})\bigotimes 
\stackrel{r_{x_{l}}-1}{\stackunder{k=0}{\sum }}|x_{l}y^{k}\func{mod}N\rangle
\langle x_{l}y^{k}\func{mod}N|$ also keep unchanged under the conditional
unitary operation in addition to the $L-N$ operator terms $\stackrel{L-1}{%
\stackunder{k_{s}=N}{\sum }}\rho _{k_{s}}$ and $\rho _{00}$, but the
contribution from all these invariant operator terms to the initial density
operator $\rho (0)$ is at most $(L-rd)/L,$ as shown in the previous section.
Then the conversion efficiency of the initial density operator $\rho (0)$
into the multiple-quantum coherences under the conditional unitary operation
will be $rd/L=(N-p-q+1)/L$ at least [14], as can be seen in the previous
section$.$ Note that $\frac{1}{2}L\leq N<L.$ The efficiency $rd/L$ is
generally not less than 50\% for a sufficient large number $N=pq$,
indicating that for an arbitrary $m\neq kr$ the initial density operator $%
\rho (0)$ is efficiently transferred into the multiple-quantum coherences of
the subensemble $S_{n}$ in the density operator $\rho (m)$ with an
efficiency generally not less than 50\% under the conditional unitary
operation $U_{I_{n_{i}}S_{n}}(y,r,N)^{m}$ on the spin ensemble $%
I_{n_{i}}S_{n}$ with a large qubit number $n$. This high conversion
efficiency will directly result in that the multiple-quantum coherences of
the subensemble $S_{n}$ in the density operator $\rho (m)$ may be detected
efficiently, as can be seen in next section. 
\[
\]
\newline
{\large 4. NMR measurement of multiple-quantum coherences}

Generally all the multiple-quantum coherences can not be observed directly
in the NMR measurement except the single quantum coherence. All
non-first-order multiple-quantum coherences usually may be detected
indirectly through the direct NMR measurement of the single-quantum
coherence. To detect multiple-quantum coherences one first needs to convert
them into single-quantum coherence by making a unitary transformation
composed of a sequence of RF pulses and interaction intervals in the spin
ensemble. In general, the density operator of the spin ensemble, for
example, $\rho (m),$ can be expanded as a quantum coherence order series
[19],

$\qquad \qquad \qquad \qquad \rho (m)=\stackrel{n}{\stackunder{p=-n}{\sum }}%
\sigma _{p}(m),\qquad \qquad \qquad \qquad \qquad \qquad \quad \ (29)$ 
\newline
where the operator $\sigma _{p}(m)$ is the $p-$order quantum coherence
operator and $\sigma _{p}(m)=\sigma _{-p}(m)^{+}$ because the density
operator is an Hermitian operator. Since the present discussion is focused
on the multiple-quantum coherences of the spin subensemble $S_{n}$ of the
spin ensemble $I_{n_{i}}S_{n}$ the order index $p$ of the expansion series
(29) in this special case is referred to the $p-$order quantum coherence of
the spin subensemble $S_{n}$. The maximum quantum order of the subensemble $%
S_{n}$ is $n.$ This means that the density operator $\rho (m)$ (29) are
generally composed of the multiple-quantum coherences with at most $2n+1$
different quantum orders range from $-n$ to $n$. The density operator $\rho
(m)$ can be detected through the spins $I$ or $S$ by applying a unitary
transformation to convert it into single quantum coherence. Here consider
the measurement method through detecting the single quantum coherence of the
spin $I$ instead of the spin $S$. Before the multiple-quantum coherences are
transferred into the single-quantum coherence of the spin $I$ they are
labelled with their own precession frequencies in order to distinguish
different order quantum coherences, and in order to observe effectively the
multiple-quantum spectral peaks of the multiple-quantum coherences one had
better label all the same order quantum coherence with a single precession
frequency. Then the spin Hamiltonian used to label the multiple-quantum
coherences of the subensemble $S_{n}$ may be chosen as

$\qquad \qquad \qquad \qquad H_{S}=\stackrel{n}{\stackunder{k=1}{\sum }}%
\omega _{Sk}S_{kz}=\omega _{S}S_{z}.\qquad \qquad \qquad \qquad \qquad \ \ \
(30)$ \newline
This labelling Hamiltonian is independent of the spins $I_{n_{i}}$. Under
this Hamiltonian the multiple-quantum spectrum of the spins $S_{n}$ has at
most $2n+1$ different order multiple-quantum peaks with their own precession
frequencies, and the frequency for all $p-$order quantum coherence is simply
equal to $p\omega _{S}$ ($p=-n,-n+1,...,n-1,n$). The frequency labelling for
the multiple-quantum coherences may be achieved by the time evolution
process of the spin ensemble ($I_{n_{i}}S_{n}$) starting at the density
operator $\rho (m)$ under the Hamiltonian (30),\newline
$\rho (m,t_{1})=\exp (-iH_{S}t_{1})\rho (m)\exp (iH_{S}t_{1})=\stackrel{n}{%
\stackunder{p=-n}{\sum }}\sigma _{p}(m)\exp (-ip\omega _{S}t_{1}).\quad \ \
(31)$ \newline
Then the density operator (31) is converted into single-quantum coherence
under the specific unitary transformation $V_{I_{n_{i}}S_{n}}(y,r,N)$ which
may be constructed through the modular exponential operation $U(y,r,N)$,

$\quad \rho _{f}(m,t_{1})=V_{I_{n_{i}}S_{n}}(y,r,N)\rho
(m,t_{1})V_{I_{n_{i}}S_{n}}(y,r,N)^{+}.\qquad \qquad \qquad \quad \ \ (32)$ 
\newline
Then the observable NMR signal for the density operator $\rho _{f}(m,t_{1})$
is given by

$S_{f}(m,t_{1})=Tr\{F\rho _{f}(m,t_{1})\}$

$\ \quad \qquad \quad
=Tr\{V_{I_{n_{i}}S_{n}}(y,r,N)^{+}FV_{I_{n_{i}}S_{n}}(y,r,N)\rho
_{f}(m,t_{1})\},\qquad \qquad \ \ (33)$ \newline
where the observable single-quantum operator $F=I_{x},$ which is
proportional to the initial density operator $\rho (0)$ (21)$.$ Because all
the same order quantum coherence has the same precession frequency the total
amplitude for a given order quantum NMR signal is the coherent sum of the
amplitudes of all the same order quantum coherence. The unitary operator $%
V_{I_{n_{i}}S_{n}}(y,r,N)$ should be constructed suitably in order that the
coherent sum of the amplitudes is constructive and the inphase
multiple-quantum spectrum is generated for all the same order quantum
coherence, otherwise the total amplitude may severely attenuate due to the
destructive coherent sum. One of the best ways to build up the unitary
operator is simple to make the unitary operator satisfy [26],

$\qquad \qquad \qquad
V_{I_{n_{i}}S_{n}}(y,r,N)=U_{I_{n_{i}}S_{n}}^{+}(y,r,N)^{m}.\qquad \qquad
\qquad \quad \ \qquad \ (34)$ \newline
This is a direct requirement of the famous time-reversal symmetry [27]. In
high-resolution NMR spectroscopy the time-reversal symmetric unitary
operator usually may not be easily created from the scalar J-coupling
interactions of a complex coupled spin system in a liquid. An alternative
method to generate the inphase multiple-quantum spectrum for all the same
order quantum coherence may be that a series of experiments are performed
starting from the same density operator, e.g., $\rho (m,t_{1})$ of Eq.(31),
and using different unitary operators $V_{I_{n_{i}}S_{n}}^{(k)}(y,r,N)$ so
that the following relation is satisfied:

$\stackrel{}{\stackunder{k}{\sum }}%
V_{I_{n_{i}}S_{n}}^{(k)}(y,r,N)^{+}FV_{I_{n_{i}}S_{n}}^{(k)}(y,r,N)=%
\varepsilon _{i}^{-1}\rho (m)^{+}=\stackrel{n}{\stackunder{p=-n}{\sum }}%
\varepsilon _{i}^{-1}\sigma _{p}(m)^{+},\quad (35)$ \newline
then the inphase multiple-quantum spectra could be generated as well by
adding coherently these experimental NMR signals, although the Hamiltonian
of every unitary operator $V_{I_{n_{i}}S_{n}}^{(k)}(y,r,N)$ may not always
satisfy the time-reversal symmetry [27]. The inphase multiple-quantum
spectrum of the density operator $\rho _{y}(m)$ of Eq.(25) may be generated
by this method later.

According to number theory [28] Euclid$^{\prime }$s algorithm may be used to
find efficiently the multiplicative inverses in modular arithmetic. The
inverse of the integer $y$ modular $N$, i.e., $y^{-1}\func{mod}N$ can be
efficiently calculated by Euclid$^{\prime }$s algorithm by taking $O((\log
_{2}N)^{3})$ steps [28]. Then according to the definition (1) of the unitary
operator $U(y,r,N)$ the inverse unitary operator $U(y,r,N)^{+}$ then can be
constructed by

$\qquad \qquad \qquad U(y,r,N)^{+}=U(y^{-1}\func{mod}N,r,N).\qquad \qquad
\qquad \qquad \quad (36)$ \newline
Then the unitary operator $U(y,r,N)^{+}$ can be efficiently constructed just
as the unitary operator $U(y,r,N)$. Consequently the inverse conditional
unitary operation $U_{I_{n_{i}}S_{n}}^{+}(y,r,N)$ can be also constructed
efficiently by the definition (14) of the conditional unitary operation $%
U_{I_{n_{i}}S_{n}}(y,r,N)$. This directly leads to that the unitary operator 
$U_{I_{n_{i}}S_{n}}^{+}(y,r,N)^{m}$ can be efficiently constructed for any
integer $m$ because $%
U_{I_{n_{i}}S_{n}}^{+}(y,r,N)^{m}=U_{I_{n_{i}}S_{n}}^{+}(y^{m},r,N).$

As an example, below the NMR multiple-quantum signal $S_{f}(m,t_{1})$ of
Eq.(33) is calculated in detailed using the unitary operator $%
V_{I_{n_{i}}S_{n}}(y,r,N)$ of Eq.(34). There are the orthogonal relations
between different order multiple-quantum coherence operators,

$\qquad \qquad \qquad Tr\{\sigma _{p}(m)^{+}\sigma _{p^{\prime
}}(m)\}=Tr\{|\sigma _{p}(m)|^{2}\}\delta _{pp^{\prime }}.\qquad \qquad
\qquad \ \ (37)$ \newline
Inserting Eq.(31) and (34) into Eq.(33) and using the orthogonal relations
(37) one can further write the NMR signal (33) as

$\quad \qquad S_{f}(m,t_{1})=\stackrel{n}{\stackunder{p=-n}{\sum }}%
\varepsilon _{i}^{-1}Tr\{|\sigma _{p}(m)|^{2}\}\exp (-ip\omega
_{S}t_{1}).\qquad \qquad \qquad \ (38)$ \newline
The multiple-quantum spectrum $S_{f}(m,\omega -p\omega _{S})$ can be
obtained by fast Fourier transforming ($t_{1}$) the time-domain NMR signal $%
S_{f}(m,t_{1})$ of Eq.(38)$.$ Then the $p-$order quantum peak in the
multiple-quantum spectrum has an intensity $I(p,m)=\varepsilon
_{i}^{-1}Tr\{|\sigma _{p}(m)|^{2}\}$ and the total intensity for the
multiple-quantum spectrum is given by

$\qquad \qquad \qquad I(m)=\stackrel{n}{\stackunder{p=-n}{\sum }}\varepsilon
_{i}^{-1}Tr\{|\sigma _{p}(m)|^{2}\}.\qquad \qquad \qquad \qquad \qquad \quad
(39)$\newline
The expansion (29) and the orthogonal relations (37) show that the total
intensity $I(m)$ also can be expressed as

$\qquad \qquad \qquad I(m)=\varepsilon _{i}^{-1}Tr\{\rho (m)^{+}\rho
(m)\}.\qquad \qquad \qquad \qquad \qquad \quad \ \ (40)$ \newline
The formula (39) and (40) are really the direct result of equation.(35), a
general method to obtain inphase multiple-quantum spectra. The total
spectral power $I(t)$ of the density operator $\rho (t)=U(t)\rho (0)U(t)^{+}$
including the contributions from both the multiple-quantum coherence and the 
$LOMSO$ operator components actually keeps unchanged when an arbitrary
unitary operation $U(t)$ is applied to the initial density operator $\rho
(0),$

$\qquad \qquad I(t)=\varepsilon ^{-1}Tr\{|\rho (t)|^{2}\}=\varepsilon
^{-1}Tr\{|\rho (0)|^{2}\}.$ \qquad \qquad \qquad \qquad $\ (41)$\newline
This property could be helpful for conveniently manipulating the modular
exponential operation. Inserting all the multiple-quantum coherence
components of the density operator $\rho (m)$ of Eq.(26) into (40) the total
intensity $I(m)$ can be calculated explicitly using the orthogonal relations
(13a) and $Tr\{I_{x,y}^{2}\}=\frac{1}{4}n_{i}2^{n_{i}}$ (the trace is only
in the subensemble $I_{n_{i}})$,

$\qquad \qquad I(m)=\frac{1}{4}n_{i}2^{n_{i}}\varepsilon _{i}\{N-\stackunder{%
j=0}{\stackrel{}{\sum }}\stackunder{k=0}{\stackrel{t-1}{\sum }}%
r_{x_{k}}\delta (m,jr_{x_{k}})\},\qquad \qquad \qquad \quad \ (42)$ \newline
where the delta function $\delta (x,y)=\delta _{xy}.$ The second term of
Eq.(42) accounts for those invariant diagonal operators $\{|x_{l}y^{k}\func{%
mod}N\rangle \langle x_{l}y^{k}\func{mod}N|\}$ under the conditional unitary
operation in Eq.(24) which have not a net contribution to the
multiple-quantum spectrum. The total intensity $I(m)$ satisfies, as shown in
previous section,

$I(m)=0$ if $m=0,r,2r,...$,

$I(m)\geq \frac{1}{4}n_{i}2^{n_{i}}\varepsilon _{i}[N-(p+q-1)]$ if $m\neq
0,r,2r,...$.\newline
If one detects only the multiple-quantum coherences of the antisymmetric
density operator $\rho _{y}(m)$ of Eq.(25) then the total intensity $%
I_{y}(m) $ of the multiple-quantum spectrum of the density operator $\rho
_{y}(m)$ can be derived as

$\qquad \qquad \qquad I_{y}(m)=\varepsilon _{i}^{-1}Tr\{\rho _{y}(m)^{+}\rho
_{y}(m)\}.\qquad \qquad \qquad \qquad \qquad \ (43)$ \newline
Since the density operator $\rho _{y}(m)$ is a pure multiple-quantum
coherence operator one can calculate the total intensity $I_{y}(m)$ directly
by inserting the density operator $\rho _{y}(m)$ of Eq.(25) into Eq.(43),

$I_{y}(m)=\varepsilon _{i}^{-1}Tr\{|\varepsilon _{i}I_{y}\bigotimes 
\stackunder{k=0}{\stackrel{t-1}{\sum }}\stackrel{r_{x_{k}}-1}{\stackunder{s=0%
}{\sum }}\sin [2\pi ms/r_{x_{k}}]|\Psi _{s}(x_{k})\rangle \langle \Psi
_{s}(x_{k})||^{2}\}$

$\qquad =\frac{1}{8}n_{i}2^{n_{i}}\varepsilon _{i}[N-\stackunder{j=0}{%
\stackrel{}{\sum }}\stackunder{k=0}{\stackrel{t-1}{\sum }}r_{x_{k}}\delta
(2m,jr_{x_{k}})],\qquad \qquad \qquad \qquad \qquad \quad \ \ (44)$ \newline
where the orthogonal relations (13a) of the eigenvectors $|\Psi
_{s}(x_{k})\rangle $ and the relations below has been used,

$\qquad \stackrel{r_{x_{k}}-1}{\stackunder{s=0}{\sum }}\exp (\pm i4\pi
ms/r_{x_{k}})=r_{x_{k}}\delta (2m,jr_{x_{k}}),$ $j=0,1,2,...\quad \qquad
\quad \ \ (45)$\newline
As can be seen in previous section, each $r_{x_{k}}$ can divide the period $%
r $ and the total number of the computational basis of those basis sets $%
S(x_{k})$ with the periods $r_{x_{k}}<r$ is $(p+q-1)$ at most including the
state $|0\rangle $ [14]$,$ that is, for any given integer $m,$ $(p+q-1)\geq 
\stackunder{j=0}{\stackrel{}{\sum }}\stackunder{k=0}{\stackrel{t-1}{\sum }}%
r_{x_{k}}\delta (2m,jr_{x_{k}})$ with the sum for index $k$ running only
over all $r_{x_{k}}<r,$ and most of the computational basis $(\geq
N-(p+q-1)) $ belong to those subsets with the period $r$. Therefore, the
total intensity $I_{y}(m)$ satisfies,

$I_{y}(m)=0$ if $m=0,r/2,r,3r/2,...,$

$I_{y}(m)\geq \frac{1}{8}n_{i}2^{n_{i}}\varepsilon _{i}[N-(p+q-1)]$ if $%
m\neq 0,r/2,r,3r/2,....$\newline
Here suppose that the period $r$ is an even integer, otherwise $%
r/2,3r/2,..., $ are not integers and hence not zero points and the density
operator $\rho _{y}(m)$ has only zero points with integer $m=0,r,2r,....$

If the initial thermal equilibrium density operator of Eq.(20) now is
transferred completely into the observable single quantum coherence, then
the generated NMR spectrum will have a total intensity $I_{0}=\varepsilon
_{i}Tr\{I_{z}^{2}\}+\stackrel{n}{\stackunder{k=1}{\sum }}\varepsilon
_{sk}Tr\{S_{kz}^{2}\}=\frac{1}{4}2^{(n+n_{i})}(n_{i}\varepsilon
_{i}+n\varepsilon _{s}).$ The conversion efficiency from the thermal
equilibrium state of Eq.(20) into the multiple-quantum coherences of the
density operators $\rho (m)$ and $\rho _{y}(m)$ under the conditional
unitary operation in Eq.(24) are given by the ratios $I_{y}(m)/I_{0}$ and $%
I_{y}(m)/I_{0},$ respectively, which satisfy,

$I(m)/I_{0}=0$ if $m=0,r,2r,...$,

$I(m)/I_{0}>n_{i}\varepsilon _{i}[N-(p+q-1)]/[2^{n}(n_{i}\varepsilon
_{i}+n\varepsilon _{s})]$ if $m\neq 0,r,2r,...$.\newline
and

$I_{y}(m)/I_{0}=0$ if $m=0,r/2,r,3r/2,...$,

$I_{y}(m)/I_{0}>\frac{1}{2}n_{i}\varepsilon
_{i}[N-(p+q-1)]/[2^{n}(n_{i}\varepsilon _{i}+n\varepsilon _{s})]$ if $m\neq
0,r/2,r,3r/2,...$.\newline
In fact, the multiple-quantum coherences of $\rho (m)$ (26) and $\rho
_{y}(m) $ (27) are generated only from the thermal equilibrium state $\rho
_{ieq}=\stackrel{n_{1}}{\stackunder{k=1}{\sum }}\varepsilon _{ik}I_{kz}$ of
the spins $I_{n_{i}}$ $($which is completely transferred into $\rho (0)$
(22)), while the thermal equilibrium state magnetization ($\rho _{seq}$) of
the spins $S_{n}$ of the initial thermal equilibrium state $\rho _{eq}$ (20)
has not a net contribution to the multiple-quantum coherences. The total
intensity for the multiple-quantum spectrum generated from the initial
density operator $\rho (0)$ (22) should be $I_{i0}=\varepsilon
_{i}Tr\{I_{z}^{2}\}=\frac{1}{4}n_{i}\varepsilon _{i}2^{(n+n_{i})}.$
Therefore, the conversion efficiency for the initial density operator (22)
into the multiple-quantum coherences $\rho _{y}(m)$ should be measured more
exactly by the ratio $I_{y}(m)/I_{i0}$ instead of the ratio $I_{y}(m)/I_{0}$%
. The ratio $I_{y}(m)/I_{i0}$ satisfies,

$I_{y}(m)/I_{i0}=0,$ $m=0,r/2,r,3r/2,...,$

$I_{y}(m)/I_{i0}\geq \frac{1}{2}[N-(p+q-1)]/2^{n},$ $m\neq 0,r/2,r,3r/2,....$%
\newline
It can be seen that the efficiency is almost independent of the qubit number 
$n$ of the spin ensemble ($I_{n_{i}}S_{n}$) for a large prime integer $N=pq$ 
$(2^{n-1}\leq N<2^{n}).$ Therefore, for a large prime integer $N$ and $m\neq
0,r/2,r,3r/2,...$ the initial thermal equilibrium density operator $\rho
_{ieq}$ is efficiently transferred into the multiple-quantum coherences $%
\rho _{y}(m)$ with an efficiency more than $1/4.$ Since there are $2n+1$
peaks in the multiple-quantum spectrum of $\rho _{y}(m)$ the intensity for
each peak of the $2n+1$ peaks, on average, is approximately inversely
proportional to the qubit number $n$, indicating that intensities for some
of the $2n+1$ multiple-quantum peaks do not reduce exponentially as the
qubit number $n$.

The density operator $\rho _{y}(m)$ of Eq.(27) is a pure multiple-quantum
coherences of the spin subensemble $S_{n}$, but $\rho (m)$ contains both the
multiple-quantum coherence and the $LOMSO$ operators. The $LOMSO$ components
may hamper the detection of zero-quantum coherence in the multiple-quantum
spectra since both have zero frequency in the multiple-quantum spectra under
the frequency labelling Hamiltonian (30). A better method may be using only
the density operator $\rho _{y}(m)$ and its intensity $I_{y}(m)$ to solve
the factoring problem. According to the scheme (35) to create inphase
multiple-quantum spectra the pulse sequence to create the output NMR signal
with the intensity $I_{y}(m)$ (43) consists of the two experiments:\newline
$\rho _{f}(m,t_{1})=U_{I_{n_{i}}S_{n}}^{+}(y^{m},r,N)\exp (-iH_{S}t_{1})\rho
_{y}(m)\exp (iH_{S}t_{1})U_{I_{n_{i}}S_{n}}(y^{m},r,N),$\newline
$\rho _{f}(m,t_{1})=U_{I_{n_{i}}S_{n}}(y^{m},r,N)R_{i}(180_{x}^{\circ })\exp
(-iH_{S}t_{1})$

$\qquad \qquad \times \rho _{y}(m)\exp (iH_{S}t_{1})R_{i}(180_{x}^{\circ
})^{+}U_{I_{n_{i}}S_{n}}^{+}(y^{m},r,N).$\newline
By adding coherently the output NMR signals of the two experiments one will
obtain the desired NMR signal with the intensity $I_{y}(m)$ (43). Since the
density operator $\rho _{y}(m)$ is obtained from the two experiments, as
shown in previous section, the complete pulse sequence to create the
intensity $I_{y}(m)$ consists of four experiments.

A more general initial density operator $\rho (0)$ is suggested below for
the modular exponential operation sequence above. For convenient treatment,
the initial density operator $\rho (0)$ still has a general $LOMSO$ operator 
$\rho _{S_{n}}(0)$ of the subensemble $S_{n}$, that is,

$\qquad \qquad \qquad \rho (0)=\alpha E+\stackrel{}{\stackunder{j=1}{\sum }}%
\rho _{I_{n_{i}}}(0)_{j}\bigotimes \rho _{S_{n}}(0)_{j}.\qquad \qquad \qquad
\qquad \ \ (46)$ \newline
The two components $\rho _{I_{n_{i}}}(0)$ and $\rho _{S_{n}}(0)$ of the
density operator $\rho (0)$ belong to the two subensembles $I_{n_{i}}$ and $%
S_{n}$ of the spin ensemble $I_{n_{i}}S_{n}$, respectively. Particularly, in
previous factoring sequence the density operator component $\rho _{S_{n}}(0)$
is the unity operator $E$, as can be seen in the initial density operator of
Eq.(21). The initial density operator can be prepared properly from the
thermal equilibrium state (20) of the spin ensemble $I_{n_{i}}S_{n}$ by a
suitable pulse sequence. Ignoring the unity operator term $\alpha E$ the
initial density operator $\rho (0)$ (46) is generally written as

$\rho (0)=\stackrel{n_{1}}{\stackunder{j=1}{\sum }}\varepsilon
_{j}I_{jx}\bigotimes \stackrel{t-1}{\stackunder{l=0}{\sum }}\stackrel{%
r_{x_{l}}-1}{\stackunder{k=0}{\sum }}\rho _{j}(x_{l}y^{k}\func{mod}%
N)|x_{l}y^{k}\func{mod}N\rangle \langle x_{l}y^{k}\func{mod}N|$

$\qquad +\stackrel{n_{1}}{\stackunder{j=1}{\sum }}\varepsilon
_{j}I_{jx}\bigotimes \stackrel{L-1}{\stackunder{k=N}{\sum }}\rho
_{j}(k)|k\rangle \langle k|$,$\quad (L=2^{n},$ $\frac{1}{2}L<N<L).\qquad
\quad \ \ (47)$\newline
where the coefficient $\rho _{j}(x_{l}y^{k}\func{mod}N)$ is the diagonal
element with the index $(x_{l}y^{k}\func{mod}N)$ of the initial density
operator component $\rho _{S_{n}}(0)_{j}$. There are the unitary
transformations according the definition (14) of the conditional modular
exponential operation:

$U_{I_{k}S_{n}}(y,r,N)^{m}I_{jx}\bigotimes |x_{l}y^{k}\func{mod}N\rangle
\langle x_{l}y^{k}\func{mod}N|U_{I_{k}S_{n}}^{+}(y,r,N)^{m}$

$=\frac{1}{2}(|1\rangle \langle 0|)_{k}\bigotimes |x_{l}y^{k+m}\func{mod}%
N\rangle \langle x_{l}y^{k}\func{mod}N|$

$+\frac{1}{2}(|0\rangle \langle 1|)_{k}\bigotimes |x_{l}y^{k}\func{mod}%
N\rangle \langle x_{l}y^{k+m}\func{mod}N|,$ if $j=k;\qquad \qquad \qquad \ \
(48a)$

$=(|1\rangle \langle 1|)_{k}\bigotimes I_{jx}\bigotimes |x_{l}y^{k+m}\func{%
mod}N\rangle \langle x_{l}y^{k+m}\func{mod}N|$

$+(|0\rangle \langle 0|)_{k}\bigotimes I_{jx}\bigotimes |x_{l}y^{k}\func{mod}%
N\rangle \langle x_{l}y^{k}\func{mod}N|,$ if $j\neq k.\qquad \qquad \quad \
\ (48b)$\newline
The unitary transformations (48a) and (48b) show that the diagonal operator $%
|x_{l}y^{k}\func{mod}N\rangle \langle x_{l}y^{k}\func{mod}N|$ of the
subensemble $S_{n}$ can not be converted into multiple-quantum coherences
but into other $LOMSO$ operators by the conditional modular exponential
operation $U_{I_{k}S_{n}}(y,r,N)^{m}$ ($k\neq j$) which Hamiltonian (15)
does not contain the operator $I_{jz}$. The unitary transformations can be
used further to calculate the time evolution of the spin ensemble with the
initial density operator $\rho (0)$ (47) under the conditional modular
exponential operation (18),

$\rho (m)=U_{I_{n_{i}}S_{n}}(y,r,N)^{m}\rho (0)U_{I_{k}S_{n}}^{+}(y,r,N)^{m}$

$=\stackrel{n_{1}}{\stackunder{j=1}{\sum }}\stackrel{t-1}{\stackunder{l=0}{%
\sum }}\stackrel{r_{x_{l}}-1}{\stackunder{k=0}{\sum }}\varepsilon _{j}\rho
_{j}(x_{l}y^{k}\func{mod}N)U_{I_{j}S_{n}}(y,r,N)^{m}$

$\times \{(E_{11})_{1}\bigotimes ...\bigotimes (E_{11})_{j-1}\bigotimes
I_{jx}\bigotimes (E_{11})_{j+1}\bigotimes ...\bigotimes (E_{11})_{n_{i}}$

$\bigotimes |x_{l}y^{k+m(n_{i}-1)}\func{mod}N\rangle \langle
x_{l}y^{k+m(n_{i}-1)}\func{mod}N|$

$+[(E_{00})_{1}\bigotimes (E_{11})_{2}\bigotimes ...\bigotimes
(E_{11})_{j-1}\bigotimes I_{jx}\bigotimes (E_{11})_{j+1}\bigotimes
...\bigotimes (E_{11})_{n_{i}}$

$+(E_{11})_{1}\bigotimes (E_{00})_{2}\bigotimes (E_{11})_{3}\bigotimes
...\bigotimes (E_{11})_{j-1}\bigotimes I_{jx}\bigotimes
(E_{11})_{j+1}\bigotimes ...$

$\bigotimes (E_{11})_{n_{i}}+...+(E_{11})_{1}\bigotimes ...\bigotimes
(E_{11})_{j-1}\bigotimes I_{jx}\bigotimes (E_{11})_{j+1}\bigotimes ...$

$\bigotimes (E_{11})_{n_{i}-1}\bigotimes (E_{00})_{n_{i}}]\bigotimes
|x_{l}y^{k+m(n_{i}-2)}\func{mod}N\rangle \langle x_{l}y^{k+m(n_{i}-2)}\func{%
mod}N|$

$+[(E_{00})_{1}\bigotimes (E_{00})_{2}\bigotimes (E_{11})_{3}\bigotimes
...\bigotimes (E_{11})_{j-1}\bigotimes I_{jx}\bigotimes
(E_{11})_{j+1}\bigotimes ...$

$\bigotimes (E_{11})_{n_{i}}+...+(E_{11})_{1}\bigotimes ...\bigotimes
(E_{11})_{j-1}\bigotimes I_{jx}\bigotimes (E_{11})_{j+1}\bigotimes
...\bigotimes (E_{11})_{n_{i}-2}$

$\bigotimes (E_{00})_{n_{i}-1}\bigotimes (E_{00})_{n_{i}}]\bigotimes
|x_{l}y^{k+m(n_{i}-3)}\func{mod}N\rangle \langle x_{l}y^{k+m(n_{i}-3)}\func{%
mod}N|$

$+...+(E_{00})_{1}\bigotimes (E_{00})_{2}\bigotimes ...\bigotimes
(E_{00})_{j-1}\bigotimes I_{jx}\bigotimes (E_{00})_{j+1}\bigotimes
...\bigotimes (E_{00})_{n_{i}}$

$\bigotimes |x_{l}y^{k}\func{mod}N\rangle \langle x_{l}y^{k}\func{mod}%
N|\}U_{I_{j}S_{n}}^{+}(y,r,N)^{m}$

$+\stackrel{n_{1}}{\stackunder{j=1}{\sum }}\varepsilon _{j}I_{jx}\bigotimes 
\stackrel{L-1}{\stackunder{k=N}{\sum }}\rho _{j}(k)|k\rangle \langle
k|,\qquad \qquad \qquad \qquad \qquad \qquad \qquad \qquad \ \ (49)$ \newline
where the operator $E_{ij}=|i\rangle \langle j|$ and the $2\times 2-$%
dimensional unity operator of the $k$th spin $I$ is expressed as $%
E_{k}^{i}=(E_{00})_{k}+(E_{11})_{k}.$ The density operator $\rho (m)$ (49)
can be simplified by using the shift-invariance identity:

$\stackrel{t-1}{\stackunder{l=0}{\sum }}\stackrel{r_{x_{l}}-1}{\stackunder{%
k=0}{\sum }}\rho _{j}(x_{l}y^{k}\func{mod}N)|x_{l}y^{k+qm}\func{mod}N\rangle
\langle x_{l}y^{k+qm}\func{mod}N|$

$=\stackrel{t-1}{\stackunder{l=0}{\sum }}\stackrel{r_{x_{l}}-1}{\stackunder{%
k=0}{\sum }}\rho _{j}(x_{l}y^{(k-qm)\func{mod}r_{x_{l}}}\func{mod}%
N)|x_{l}y^{k}\func{mod}N\rangle \langle x_{l}y^{k}\func{mod}N|,\qquad \ \
(50)$ \newline
where $0\leq (k-qm)\func{mod}r_{x_{l}}<r_{x_{l}}$ for any integers $q$ and $%
m $. For example, one of those operator terms in the density operator $\rho
(m) $ (49) is calculated in detailed below,

$\rho ^{q}(m)=U_{I_{j}S_{n}}(y,r,N)^{m}(E_{11})_{1}\bigotimes ...\bigotimes
(E_{11})_{j-1}\bigotimes I_{jx}\bigotimes (E_{11})_{j+1}$

$\bigotimes ...\bigotimes (E_{11})_{n_{i}}\bigotimes \stackrel{t-1}{%
\stackunder{l=0}{\sum }}\stackrel{r_{x_{l}}-1}{\stackunder{k=0}{\sum }}%
\varepsilon _{j}\rho _{j}(x_{l}y^{k}\func{mod}N)$

$\times |x_{l}y^{k+qm}\func{mod}N\rangle \langle x_{l}y^{k+qm}\func{mod}%
N|U_{I_{j}S_{n}}^{+}(y,r,N)^{m}$

$=\frac{1}{2}(E_{11})_{1}\bigotimes ...\bigotimes (E_{11})_{j-1}\bigotimes
(|1\rangle \langle 0|)_{j}\bigotimes (E_{11})_{j+1}\bigotimes ...\bigotimes
(E_{11})_{n_{i}}$

$\bigotimes \stackrel{t-1}{\stackunder{l=0}{\sum }}\stackrel{r_{x_{l}}-1}{%
\stackunder{k=0}{\sum }}\varepsilon _{j}\rho _{j}(x_{l}y^{[k-qm]\func{mod}%
r_{x_{l}}}\func{mod}N)|x_{l}y^{k+m}\func{mod}N\rangle \langle x_{l}y^{k}%
\func{mod}N|$

$+\frac{1}{2}(E_{11})_{1}\bigotimes ...\bigotimes (E_{11})_{j-1}\bigotimes
(|0\rangle \langle 1|)_{j}\bigotimes (E_{11})_{j+1}\bigotimes ...\bigotimes
(E_{11})_{n_{i}}$

$\bigotimes \stackrel{t-1}{\stackunder{l=0}{\sum }}\stackrel{r_{x_{l}}-1}{%
\stackunder{k=0}{\sum }}\varepsilon _{j}\rho _{j}(x_{l}y^{[k-qm]\func{mod}%
r_{x_{l}}}\func{mod}N)$

$\times |x_{l}y^{k}\func{mod}N\rangle \langle x_{l}y^{k+m}\func{mod}N|,$ $%
(q=n_{i}-1).\qquad \qquad \qquad \qquad \qquad \ \ (51)$\newline
One then further calculates the contribution of the operator term (51) to
the total intensity $I(m)$ of the multiple-quantum spectrum of the
subensemble $S_{n}$ of the density operator $\rho (m)$ (49)$.$ When $m\neq
k^{\prime }r_{x_{l}},$ $(k^{\prime }=0,1,...;l=0,1,...,t-1)$, the
contribution is given by

$2\varepsilon _{i}\stackrel{t-1}{\stackunder{l=1}{\sum }}\stackrel{%
r_{x_{l}}-1}{\stackunder{k=0}{\sum }}|\frac{1}{2}\rho _{j}(x_{l}y^{[k-qm]%
\func{mod}r_{x_{l}}}\func{mod}N)|^{2}=\frac{1}{2}\varepsilon _{i}\stackrel{%
N-1}{\stackunder{k=1}{\sum }}\rho _{j}(k)^{2}.$\newline
where the diagonal operator term $\rho _{00}=|0\rangle \langle 0|$ is not
included. When $m=k^{\prime }r$ the contribution is nothing to the
multiple-quantum spectrum, that is, the integers $m=k^{\prime }r$ are the
zero points of the multiple-quantum spectrum. When $m=k^{\prime }r_{x_{l}}$
but $m\neq k^{\prime }r$ the contribution of the term (51) satisfies

$2\varepsilon _{i}\stackrel{t-1}{\stackunder{l=1}{\sum }}\stackrel{%
r_{x_{l}}-1}{\stackunder{k=0}{\sum }}|\frac{1}{2}\rho _{j}(x_{l}y^{[k-qm]%
\func{mod}r_{x_{l}}}\func{mod}N)|^{2}$

$\qquad \geq \frac{1}{2}\varepsilon _{i}\{\stackrel{N-1}{\stackunder{k=0}{%
\sum }}\rho _{j}(k)^{2}-(p+q-1)|\rho _{j}(k)^{2}|_{\max }\}$\newline
where $|\rho _{j}(k)|_{\max }$ is the maximum diagonal element of the
initial density operator component $\rho _{S_{n}}(0)_{j}.$ Now the total
intensity $I(m)$ of the multiple-quantum spectrum can be calculated from the
density operator (49) using the orthogonal relations between any pair of
operator terms including the operator term (51) in the density operator $%
\rho (m)$ (49). When $m\neq k^{\prime }r_{x_{l}},$ the total intensity $I(m)$
is given by

$I(m)=\frac{1}{2}2^{n_{i}-1}\varepsilon _{i}\stackrel{n_{1}}{\stackunder{j=1%
}{\sum }}\stackrel{N-1}{\stackunder{k=1}{\sum }}\rho _{j}(k)^{2}.$\newline
When $m=k^{\prime }r$ the total intensity is zero, that is,

$I(m)=0$ if $m=0,r,2r,...,$\newline
indicating that the integers $m=k^{\prime }r$ are the zero points of the
multiple-quantum spectrum. When $m=k^{\prime }r_{x_{l}}$ but $m\neq
k^{\prime }r$ the total intensity $I(m)$ satisfies

$I(m)\geq \frac{1}{2}2^{n_{i}-1}\varepsilon _{i}\stackrel{n_{1}}{\stackunder{%
j=1}{\sum }}\{\stackrel{N-1}{\stackunder{k=0}{\sum }}\rho
_{j}(k)^{2}-(p+q-1)|\rho _{j}(k)^{2}|_{\max }\}$.\newline
Obviously, the initial density operator component $\rho _{S_{n}}(0)_{j}$ of
the initial density operator $\rho (0)$ (47) can be efficiently transferred
into the multiple-quantum coherences under the conditional unitary operation 
$U_{I_{n_{i}}S_{n}}(y,r,N)^{m}$ with any integer $m$ except the zero points $%
m=k^{\prime }r$ when the density operator component $\rho _{S_{n}}(0)_{j}$
satisfies,

$\qquad \qquad \stackrel{N-1}{\stackunder{k=0}{\sum }}\rho
_{j}(k)^{2}>>(p+q-1)|\rho _{j}(k)^{2}|_{\max }.\qquad \qquad \qquad \qquad
\quad \ (52a)$ \newline
and

$\qquad \qquad \{\stackrel{N-1}{\stackunder{k=0}{\sum }}\rho
_{j}(k)^{2}\}^{-1}\{\stackrel{L-1}{\stackunder{k=0}{\sum }}\rho
_{j}(k)^{2}\}\thicksim poly(n)$\qquad \qquad \quad \qquad \qquad $\ (52b)$ 
\newline
Using the auxiliary experiment $\rho
(-m)=U_{I_{n_{i}}S_{n}}^{+}(y,r,N)^{m}\rho (0)U_{I_{n_{i}}S_{n}}(y,r,N)^{m}$
one can separate the antisymmetric part $\rho _{y}(m)$ $(\rho _{y}(m)=\frac{1%
}{2}\rho (m)-\frac{1}{2}\rho (-m))$ from the density operator $\rho (m)$
(49). The density operator $\rho _{y}(m)$ is a pure multiple-quantum
coherence operator. For example, the operator term (51) contains a component
of the density operator $\rho _{y}(m)$:

$\rho _{y}^{q}(m)=\frac{1}{4}(E_{11})_{1}\bigotimes ...\bigotimes
(E_{11})_{j-1}\bigotimes I_{jx}\bigotimes (E_{11})_{j+1}\bigotimes
...\bigotimes (E_{11})_{n_{i}}$

$\bigotimes \stackrel{t-1}{\stackunder{l=0}{\sum }}\stackrel{r_{x_{l}}-1}{%
\stackunder{k=0}{\sum }}\varepsilon _{j}[\rho _{j}(x_{l}y^{[k-qm]\func{mod}%
r_{x_{l}}}\func{mod}N)-\rho _{j}(x_{l}y^{[k+qm+m]\func{mod}r_{x_{l}}}\func{%
mod}N)]$

$\times \{|x_{l}y^{k+m}\func{mod}N\rangle \langle x_{l}y^{k}\func{mod}%
N|+|x_{l}y^{k}\func{mod}N\rangle \langle x_{l}y^{k+m}\func{mod}N|\}$

$+\frac{1}{4}(E_{11})_{1}\bigotimes ...\bigotimes (E_{11})_{j-1}\bigotimes
I_{jy}\bigotimes (E_{11})_{j+1}\bigotimes ...\bigotimes (E_{11})_{n_{i}}$

$\bigotimes \stackrel{t-1}{\stackunder{l=0}{\sum }}\stackrel{r_{x_{l}}-1}{%
\stackunder{k=0}{\sum }}\varepsilon _{j}[\rho _{j}(x_{l}y^{[k-qm]\func{mod}%
r_{x_{l}}}\func{mod}N)$

$+\rho _{j}(x_{l}y^{[k+qm+m]\func{mod}r_{x_{l}}}\func{mod}N)]\{i|x_{l}y^{k}%
\func{mod}N\rangle \langle x_{l}y^{k+m}\func{mod}N|$

$-i|x_{l}y^{k+m}\func{mod}N\rangle \langle x_{l}y^{k}\func{mod}N|\},$ $%
(q=n_{i}-1).\qquad \qquad \qquad \qquad \qquad (53)$\newline
One sees that the zero points of the density operator $\rho _{y}(m)$ are $%
m=k^{\prime }r/2,$ that is, $\rho _{y}(k^{\prime }r/2)=0.$ When $m\neq
k^{\prime }r_{x_{l}}/2,$ it is easy to calculate the contribution of the
operator term $\rho _{y}^{q}(m)$ (53) to the total intensity $I_{y}(m)$ of
the multiple-quantum spectrum of the density operator $\rho _{y}(m),$

$I_{y}^{q}(m)=Tr\{|\rho _{y}^{q}(m)|^{2}\}=\frac{1}{4}\varepsilon _{i}%
\stackrel{N-1}{\stackunder{k=1}{\sum }}\rho _{j}(k)^{2}.$\newline
When $m=k^{\prime }r_{x_{l}}/2$ but $m\neq k^{\prime }r/2$ the contribution
of the operator term (53) satisfies, $I_{y}^{q}(m)\geq \frac{1}{4}%
\varepsilon _{i}\{\stackrel{N-1}{\stackunder{k=1}{\sum }}\rho
_{j}(k)^{2}-(p+q-1)|\rho _{j}(k)^{2}|_{\max }\}.$ The total intensity $%
I_{y}(m)$ then can be calculated through the density operator $\rho _{y}(m)$%
. It is easy to prove that the intensity $I_{y}(m)$ satisfies,\newline
$I_{y}(m)=0$ if $m=0,r/2,r,3r/2,...;$\newline
$I_{y}(m)\geq \frac{1}{4}\varepsilon _{i}2^{n_{i}-1}\{\stackrel{N-1}{%
\stackunder{k=0}{\sum }}\rho _{j}(k)^{2}-(p+q-1)|\rho _{j}(k)^{2}|_{\max }\}$
if $m\neq 0,r/2,r,3r/2,....$\newline
Therefore, the pure multiple-quantum coherences of the density operator $%
\rho _{y}(m)$ are efficiently created by the conditional modular exponential
operation $U_{I_{n_{i}}S_{n}}(y,r,N)^{m}$ with any integer $m$ except the
zero points $m=k^{\prime }r/2$ when the initial density operator component $%
\rho _{S_{n}}(0)_{j}$ satisfies the conditions (52a) and (52b).

In order to calculate analytically the multiple-quantum spectrum of the
density operators $\rho (m)$ and $\rho _{y}(m)$ a compact and analytical
derivation for the density operators is given below. With the help of the
Fourier transform (12b) and the Hermitian property of the density operator
the initial density operator $\rho (0)$ (47) can be written in a symmetrical
form

$\rho (0)=\stackrel{n_{1}}{\stackunder{j=1}{\sum }}\varepsilon
_{j}I_{jx}\bigotimes \stackunder{l=0}{\stackrel{t-1}{\sum }}\stackrel{%
r_{x_{l}}-1}{\stackunder{s=0}{\sum }}\stackrel{r_{x_{l}}-1}{\stackunder{%
s^{\prime }=0}{\sum }}\frac{1}{2}\{\rho _{j}(s-s^{\prime },x_{l})|\Psi
_{s}(x_{l})\rangle \langle \Psi _{s^{\prime }}(x_{l})|$

$+\rho _{j}(s^{\prime }-s,x_{l})|\Psi _{s^{\prime }}(x_{l})\rangle \langle
\Psi _{s}(x_{l})|\}+\stackrel{n_{1}}{\stackunder{j=1}{\sum }}\varepsilon
_{j}I_{jx}\bigotimes \stackrel{L-1}{\stackunder{k=N}{\sum }}\rho
_{j}(k)|k\rangle \langle k|\qquad \quad \ \ (54)$ \newline
where the coefficient $\rho _{j}(s-s^{\prime },x_{l})$ is given by

$\rho _{j}(s-s^{\prime },x_{l})=\frac{1}{r_{x_{l}}}\stackrel{r_{x_{l}}-1}{%
\stackunder{k=0}{\sum }}\rho _{j}(x_{l}y^{k}\func{mod}N)\exp [-i2\pi
k(s-s^{\prime })/r_{x_{l}}].\qquad \ \ (55)$ \newline
Now the time evolution of the spin ensemble is calculated with the aid of
the Hamiltonian of Eq.(15) and the eigen-equation (4a) when acting the
conditional unitary operation $U_{I_{n_{i}}S_{n}}(y,r,N)^{m}$ on the density
operator $\rho (0)$ of Eq.(54),

$\rho (m)=U_{I_{n_{i}}S_{n}}(y,r,N)^{m}\rho
(0)U_{I_{n_{i}}S_{n}}^{+}(y,r,N)^{m}$

$=\frac{1}{2}\stackunder{l=0}{\stackrel{t-1}{\sum }}\stackrel{r_{x_{l}}-1}{%
\stackunder{s=0}{\sum }}\stackrel{r_{x_{l}}-1}{\stackunder{s^{\prime }=0}{%
\sum }}\exp [-i\pi n_{i}m(s-s^{\prime })/r_{x_{l}}]\exp [i2\pi m(s-s^{\prime
})/r_{x_{l}}I_{z}]$

$\times \{\stackrel{n_{1}}{\stackunder{j=1}{\sum }}\varepsilon _{j}\rho
_{j}(s-s^{\prime },x_{l})(I_{jx}\cos [2\pi ms^{\prime
}/r_{x_{l}}]-I_{jy}\sin [2\pi ms^{\prime }/r_{x_{l}}])\}$

$\bigotimes |\Psi _{s}(x_{l})\rangle \langle \Psi _{s^{\prime }}(x_{l})|+%
\frac{1}{2}\stackunder{l=0}{\stackrel{t-1}{\sum }}\stackrel{r_{x_{l}}-1}{%
\stackunder{s=0}{\sum }}\stackrel{r_{x_{l}}-1}{\stackunder{s^{\prime }=0}{%
\sum }}\exp [-i\pi n_{i}m(s^{\prime }-s)/r_{x_{l}}]$

$\times \{\stackrel{n_{1}}{\stackunder{j=1}{\sum }}\varepsilon _{j}\rho
_{j}(s^{\prime }-s,x_{l})(I_{jx}\cos [2\pi ms^{\prime
}/r_{x_{l}}]-I_{jy}\sin [2\pi ms^{\prime }/r_{x_{l}}])\}$

$\times \exp [i2\pi m(s^{\prime }-s)/r_{x_{l}}I_{z}]\bigotimes |\Psi
_{s^{\prime }}(x_{l})\rangle \langle \Psi _{s}(x_{l})|$

$+\stackrel{n_{1}}{\stackunder{j=1}{\sum }}\varepsilon _{j}I_{jx}\bigotimes 
\stackrel{L-1}{\stackunder{k=N}{\sum }}\rho _{j}(k)|k\rangle \langle
k|.\qquad \qquad \qquad \qquad \qquad \qquad \qquad \qquad \quad (56)$%
\newline
The unitary diagonal operator $\exp [i2\pi m(s-s^{\prime })/r_{x_{l}}I_{z}]$
can be expanded in the $LOMSO$ subspace [29] of the subensemble $I_{n_{i}}$,

$\exp [i2\pi m(s-s^{\prime })/r_{x_{l}}I_{z}]=\alpha _{0}F_{0}+\alpha
_{1}F_{1}+\alpha _{2}F_{2}+...+\alpha _{n_{i}}F_{n_{i}}\qquad \qquad (57)$%
\newline
where the operator $F_{k}$ is the full symmetrical $k-$body interaction
basis operator of the $LOMSO$ subspace [21, 29],

$F_{0}=E,$ $F_{1}=I_{z},$ $F_{2}=\stackrel{n_{i}}{\stackunder{l>k=1}{\sum }}%
2I_{kz}I_{lz},$ $F_{3}=\stackrel{n_{i}}{\stackunder{m>l>k=1}{\sum }}%
4I_{kz}I_{lz}I_{mz},...,$

$F_{n_{i}}=2^{n_{i}-1}I_{1z}I_{2z}...I_{n_{i}z},$\newline
and the coefficient $\alpha _{p}$ is generally expressed as

$\alpha _{p}=\stackrel{n_{i}}{\stackunder{q=0}{\sum }}c_{pq}\exp [i2\pi
m(s-s^{\prime })/r_{x_{l}}(n_{i}/2-q)],\qquad \qquad \qquad \qquad \qquad
\quad \ \ (58)$ \newline
where the real coefficient $c_{pq}$ can be determined using the method in
Ref.[29]. By inserting Eqs.(55), (57), and (58) into Eq.(56) and then using
the Fourier transform (12a) and dividing the $LOMSO$ operator $F_{p}$ into
two parts: $F_{p}=F_{1p}^{j}+2I_{jz}F_{2p}^{j}$ the density operator $\rho
(m)$ can be expressed in terms of the conventional computational basis,

$\rho (m)=-\frac{1}{2}\stackunder{l=0}{\stackrel{t-1}{\sum }}$ $\stackrel{%
r_{x_{l}}-1}{\stackunder{k=0}{\sum }}\stackrel{n_{1}}{\stackunder{p,q=0}{%
\sum }}\stackrel{n_{1}}{\stackunder{j=1}{\sum }}c_{pq}\varepsilon
_{j}I_{jy}Q_{1}(j,p,x_{l},k,m)$

$\bigotimes (\frac{1}{2i})\{+|x_{l}y^{k+qm}\func{mod}N\rangle \langle
x_{l}y^{k+qm+m}\func{mod}N|$

$\qquad -|x_{l}y^{k+qm+m}\func{mod}N\rangle \langle x_{l}y^{k+qm}\func{mod}%
N|\}$

$+\frac{1}{2}\stackunder{l=0}{\stackrel{t-1}{\sum }}$ $\stackrel{r_{x_{l}}-1%
}{\stackunder{k=0}{\sum }}\stackrel{n_{1}}{\stackunder{p,q=0}{\sum }}%
\stackrel{n_{1}}{\stackunder{j=1}{\sum }}c_{pq}\varepsilon
_{j}I_{jx}Q_{1}(j,p,x_{l},k,m)$

$\bigotimes (\frac{1}{2})\{+|x_{l}y^{k+qm}\func{mod}N\rangle \langle
x_{l}y^{k+qm+m}\func{mod}N|$

$\qquad +|x_{l}y^{k+qm+m}\func{mod}N\rangle \langle x_{l}y^{k+qm}\func{mod}%
N|\}$

$+\stackrel{n_{1}}{\stackunder{j=1}{\sum }}\varepsilon _{j}I_{jx}\bigotimes 
\stackrel{L-1}{\stackunder{k=N}{\sum }}\rho _{j}(k)|k\rangle \langle
k|,\qquad \qquad \qquad \qquad \qquad \qquad \qquad \qquad \ \ (59)$ \newline
where the operator function $Q_{1}(j,p,x_{l},k,m)$ is defined as

$Q_{1}(j,p,x_{l},k,m)=F_{1p}^{j}[\rho _{j}(x_{l}y^{k}\func{mod}N)+\rho
_{j}(x_{l}y^{k+m}\func{mod}N)]$

$\qquad \qquad \qquad \qquad +F_{2p}^{j}[\rho _{j}(x_{l}y^{k}\func{mod}%
N)-\rho _{j}(x_{l}y^{k+m}\func{mod}N)]$\newline
Using the auxiliary experiment $\rho
(-m)=U_{I_{n_{i}}S_{n}}^{+}(y,r,N)^{m}\rho (0)U_{I_{n_{i}}S_{n}}(y,r,N)^{m}$
one can further separate the antisymmetric part $\rho _{y}(m)$ from the
density operator $\rho (m)$, $(\rho _{y}(m)=\frac{1}{2}\rho (m)-\frac{1}{2}%
\rho (-m)),$

$\rho _{y}(m)=-\frac{1}{4}\stackunder{l=0}{\stackrel{t-1}{\sum }}$ $%
\stackrel{r_{x_{l}}-1}{\stackunder{k=0}{\sum }}\stackrel{n_{1}}{\stackunder{%
p,q=0}{\sum }}\stackrel{n_{1}}{\stackunder{j=1}{\sum }}c_{pq}\varepsilon
_{j}I_{jy}Q_{1}(j,p,x_{l},k,m)$

$\bigotimes (\frac{1}{2i})\{+|x_{l}y^{k+qm}\func{mod}N\rangle \langle
x_{l}y^{k+qm+m}\func{mod}N|$

$\qquad -|x_{l}y^{k+qm+m}\func{mod}N\rangle \langle x_{l}y^{k+qm}\func{mod}%
N|\}$

$-\frac{1}{4}\stackunder{l=0}{\stackrel{t-1}{\sum }}$ $\stackrel{r_{x_{l}}-1%
}{\stackunder{k=0}{\sum }}\stackrel{n_{1}}{\stackunder{p,q=0}{\sum }}%
\stackrel{n_{1}}{\stackunder{j=1}{\sum }}c_{pq}\varepsilon
_{j}I_{jy}Q_{2}(j,p,x_{l},k+qm,m)$

$\bigotimes (\frac{1}{2i})\{+|x_{l}y^{k}\func{mod}N\rangle \langle
x_{l}y^{k+m}\func{mod}N|$

$\qquad -|x_{l}y^{k+m}\func{mod}N\rangle \langle x_{l}y^{k}\func{mod}N|\}$

$+\frac{1}{4}\stackunder{l=0}{\stackrel{t-1}{\sum }}$ $\stackrel{r_{x_{l}}-1%
}{\stackunder{k=0}{\sum }}\stackrel{n_{1}}{\stackunder{p,q=0}{\sum }}%
\stackrel{n_{1}}{\stackunder{j=1}{\sum }}c_{pq}\varepsilon
_{j}I_{jx}Q_{1}(j,p,x_{l},k,m)$

$\bigotimes (\frac{1}{2})\{+|x_{l}y^{k+qm}\func{mod}N\rangle \langle
x_{l}y^{k+qm+m}\func{mod}N|$

$\qquad +|x_{l}y^{k+qm+m}\func{mod}N\rangle \langle x_{l}y^{k+qm}\func{mod}%
N|\}$

$-\frac{1}{4}\stackunder{l=0}{\stackrel{t-1}{\sum }}$ $\stackrel{r_{x_{l}}-1%
}{\stackunder{k=0}{\sum }}\stackrel{n_{1}}{\stackunder{p,q=0}{\sum }}%
\stackrel{n_{1}}{\stackunder{j=1}{\sum }}c_{pq}\varepsilon
_{j}I_{jx}Q_{2}(j,p,x_{l},k+qm,m)$

$\bigotimes (\frac{1}{2})\{+|x_{l}y^{k}\func{mod}N\rangle \langle
x_{l}y^{k+m}\func{mod}N|$

$\qquad +|x_{l}y^{k+m}\func{mod}N\rangle \langle x_{l}y^{k}\func{mod}%
N|\}\qquad \qquad \qquad \qquad \qquad \qquad \qquad \ (60)$ \newline
where the operator function $Q_{2}(j,p,x_{l},k,m)$ is defined as

$Q_{2}(j,p,x_{l},k,m)=F_{1p}^{j}[\rho _{j}(x_{l}y^{k}\func{mod}N)+\rho
_{j}(x_{l}y^{k+m}\func{mod}N)]$

$\qquad \qquad \qquad \qquad -F_{2p}^{j}[\rho _{j}(x_{l}y^{k}\func{mod}%
N)-\rho _{j}(x_{l}y^{k+m}\func{mod}N)]$\newline
Obviously, the density operator $\rho _{y}(m)$ is antisymmetric, that is, $%
\rho _{y}(kr/2+m)=-\rho _{y}(kr/2-m)$ ($k=0,1,...$)$,$ and it is a pure
multiple-quantum coherence operator of the subensemble $S_{n}.$ The
antisymmetric property might be helpful for speeding up the searching for
the zero points of the density operator $\rho _{y}(m)$. 
\[
\]
\newline
{\large 5. Searching for the period of modular exponential function}

In previous sections it has been shown that the modular exponential
operation can be performed easily on an NMR quantum computer just like on a
classical digital computer. The classical computer outputs the value of the
modular exponential function $f(y,m,N)=y^{m}\func{mod}N$ given the input
integers $y,$ $m,$ and $N$, while the NMR quantum computer outputs the
multiple-quantum spectrum of the spin ensemble which intensity does not
reduce exponentially as the qubit number of the spin ensemble. Both the
classical and quantum computations of the modular exponential function have
the same computational complexity. In classical computation the values of
the modular exponential function with different inputs, i.e., the integer $%
m, $ are generally independent on each other. Therefore it is a hard problem
to find the period $r$ of the modular exponential function on a classical
computer. However, the essential difference for the quantum computer from
the classical one is that the computational process on the quantum computer
obeys the unitary dynamics of quantum mechanics. Then in the factoring
problem the quantum computational process and output (through the density
operator) are governed by the Liouville-von Neumann equation or the Schr$%
\ddot{o}$dinger equation where the integer $m$ acts as the discrete time
variable, as can be seen below, and therefore the output results at
different times ($m$) really correlate to each other. This essential point
could form the base to solve efficiently the factoring problem and play a
key important role for the quantum computer outperforming the classical one
in solving the factoring problem.

In general, the amplitudes and phases of multiple-quantum coherences of $%
\rho (m)$ and $\rho _{y}(m)$ with different quantum orders are dependent on
the integer $m$ in the spin ensemble, which are described by the
Liouville-von Neumann equation. In particular, the amplitude and phase for
the long-range-interaction and higher-order multiple-quantum coherences
could be helpful for efficiently searching for the period $r$. The
Liouville-von Neumann equation with Hamiltonian $H(y,r,N)$ that governs the
unitary dynamical process during the modular exponential operation (1) in a
spin ensemble can be written as

$\qquad \qquad d\rho (t)/dt=-i[H(y,r,N),\rho (t)],$ $(\hslash =1).\qquad
\qquad \qquad \qquad \ \ (61)$\newline
The solution to the Liouville equation then is given formally by

$\qquad \qquad \rho (t)=U(t)\rho (0)U(t)^{+}\qquad \qquad \qquad \qquad
\qquad \qquad \qquad \qquad \ \ (62)$ \newline
where the propagator is written as

$\qquad \qquad U(t)=\exp [-itH(y,r,N)].\qquad \qquad \qquad \qquad \qquad
\qquad \qquad \ (63)$ \newline
By comparing the unitary operator $U(t)$ with the modular exponential
operator $U(y,r,N)^{m}$ of Eq.(7)\ one sees that the integer $m$ in the
modular exponential operator is really equivalent to the time variable $t$
and their difference is merely that the time variable $t$ is continuous but
the integer $m$ discrete. Therefore, the propagator is time periodic: $%
U(t)=U(t+r),$ where the period $r$ needs to be determined in the factoring
problem. Below it is assumed that the Hamiltonian $H(y,r,N)$ consists of a
dominating and a relative small operator components,

$\qquad \qquad H(y,r,N)=H_{0}+H_{1}.\qquad \qquad \qquad \qquad \qquad
\qquad \qquad \qquad \ (64)$ \newline
The dominating term $H_{0}$ is a specific order quantum operator, for
example, a zero-quantum coherence operator, while the small operator term $%
H_{1}$ is usually a multiple-quantum operator. By making the coordinate
frame transformation: $\rho _{r}(t)=\exp (iH_{0}t)\rho (t)\exp (-iH_{0}t),$
here the frame is called the interaction frame defined by the Hamiltonian $%
H_{0},$ the Liouville equation (61) is rewritten as

$\qquad \qquad d\rho _{r}(t)/dt=-i[\hat{H}_{1}(t),\rho _{r}(t)],\qquad
\qquad \qquad \qquad \qquad \qquad \quad \ \ (65)$ \newline
with the time-dependent Hamiltonian in the interaction frame:

$\qquad \qquad \hat{H}_{1}(t)=\exp (iH_{0}t)H_{1}\exp (-iH_{0}t).\qquad
\qquad \qquad \qquad \qquad \quad \ (66)$ \newline
The solution to the Liouville equation (66) is given in form

$\qquad \qquad \rho _{r}(t)=U_{1}(t,t_{0})\rho
(t_{0})U_{1}(t,t_{0})^{+}\qquad \qquad \qquad \qquad \qquad \qquad \quad
(67) $ \newline
with the propagator in the interaction frame:

$\qquad \qquad U_{1}(t,t_{0})=T\exp (-i\stackrel{t}{\stackunder{t_{0}}{\int }%
}\hat{H}_{1}(t^{\prime })dt^{\prime }),\qquad \qquad \qquad \qquad \qquad
\quad \ (68)$ \newline
where the operator $T$ is Dyson time-ordering operator. To see more clearly
the time evolution process in the interaction frame the solution of Eq.(67)
is expanded,

$\rho (t)=\exp (-iH_{0}t)\{\rho (0)-it[\hat{H}_{1}(t),\rho (0)]$

$\qquad \qquad -\frac{1}{2}t^{2}[\hat{H}_{1}(t),[\hat{H}_{1}(t),\rho
(0)]]+...\}\exp (iH_{0}t).\qquad \qquad \qquad \quad \ \ \ (69)$ \newline
Suppose that the initial density operator $\rho (0)$ is a $LOMSO$ operator,
for example, the initial density operator component $\rho _{S_{n}}(0)$ of
Eq.(46) that is a $LOMSO$ operator of the subensemble $S_{n}$. It can be
seen clearly from the expansion (69) how the initial density operator is
converted into multiple-quantum coherences as time development. Since the
zero-quantum Hamiltonian $H_{0}$ is dominating the initial $LOMSO$ density
operator is converted efficiently into the zero-quantum coherence at a short
time, e.g., $t=1$ ($m=1$). In general, the $p-$order quantum peak is
strongest at a short time (a small $m$) if the Hamiltonian $H(y,r,N)$
contains a dominating $p-$order quantum coherence operator. This ensures
that the NMR multiple-quantum signals at the time points $t=0$ and $t=1$ can
be precisely distinguished experimentally without an exponential resource by
the multiple-quantum spectroscopic method. The measurement precision is
important on a quantum computer, while it is not any problem in a classical
computer. Since the multiple-quantum Hamiltonian $\hat{H}_{1}(t)$ is small
the nonzero-order multiple-quantum coherences will grow slowly and
monotonously in a long time interval. At the same time the zero-quantum
coherence first increases quickly and reaches its maximum and then decreases
gradually as time development because part of the initial density operator
is converted into the multiple-quantum coherences. The same time evolution
behavior of the density operator $\rho (t)$ near the zero point $\rho (0)$
also occurs at other zero points $\rho (kr)$ ($k=0,1,...$) due to the period
of the density operator, $\rho (kr)=$ $\rho (0).$ If such time development
behavior for the zero-quantum and nonzero-order quantum coherences continues
in a time interval $\Delta T$ satisfying $r/\Delta T\thicksim poly(n)$ then
the searching for the zero points of the multiple-quantum spectra will be
polynomial-time on the NMR quantum computer. The searching efficiency is
proportional to the time interval $\Delta T$, that is, the longer the time
interval $\Delta T$ the higher the efficiency. The searching efficiency will
decrease if the density operator $\rho (t)$ arrives at its steady state at a
shorter time. Here the steady state implies that the intensity $I_{y}(p,t)$
for any $p-$order quantum coherence does not change as the time. The
situation may occur when the multiple-quantum Hamiltonian $H_{1}$ is not
small, however, even in this case the searching is still efficient if the
initial integer $m$ is sufficiently near the zero points. Obviously, the
searching for the zero points is locally efficient in a small region near
the zero points. The steady-state problem is harmful for the present
factoring algorithm to find efficiently the zero points in a spin ensemble
and needs to be overcome. It is closely related to the Hamiltonian $H(y,r,N)$
and the distribution of different order quantum transitions in a spin
ensemble. It is possible to overcome the steady-state problem by
manipulating the Hamiltonian $H(y,r,N)$ and choosing the proper initial
density operator in the factoring sequence.

The distribution of different order quantum transitions for a spin ensemble
with $n$ non-equivalent spins-1/2 has been found [19]. The number of the
zero-quantum transitions is $Z_{0}=\frac{1}{2}\{( 
\begin{array}{l}
2n \\ 
n
\end{array}
)-2^{n}\}$ and for $p-$order quantum transitions $Z_{p}=( 
\begin{array}{l}
2n \\ 
n-p
\end{array}
),$ $p=1,2,...,n.$ In the case of the large $n$ and relative small $p\neq 0$
the $p-$order quantum transition number can be approximated by Stirling
formulae, $Z_{p}=4^{n}(\pi n)^{-1/2}\exp (-p^{2}/n).$ This indicates that
the population distribution of different order quantum transitions is
extremely nonuniform in the spin ensemble. In general, the lower order
quantum transitions such as zero-, single-, and double-quantum transitions
are much more than those higher-order quantum transitions in a spin
ensemble. The spectral intensity of the $p-$order quantum transition is
generally proportional to the probability of the $p-$order quantum
transition in the distribution, although this is not absolute. Therefore, it
is better to choose lower order quantum transition spectral peaks such as
zero-, single- or double-quantum transitions to help the searching for the
period $r\ $in the factoring algorithm.

It has been shown in previous sections that the total conversion efficiency $%
I_{y}(m)/I_{i0}$ of the multiple-quantum coherences is almost independent of
the qubit number $n$ when $m$ is not a zero point and both the conditions
(52a) and (52b) are met. Because there are only $(2n+1)$ spectral peaks in
the multiple-quantum spectrum, on average, each peak intensity is
approximately inversely proportional to the qubit number $n$ even when the
integer $m=1.$ Then there are at least some peaks among the $2n+1$ peaks,
for example, the $p-$order multiple-quantum peak, which intensity $%
I_{y}(p,m) $ can be detected precisely without an exponential resource.
Consequently, with the factoring sequence in previous sections one can fix
experimentally the zero points $m_{0}=kr/2$ ($k=0,1,...,$) from a small
neighbor region ($m_{0}\pm 1$ at least) of the zero points without an
exponential resource, and thus the factoring sequence is locally efficient
in a small neighbor region of the zero points at least. However, it is not
clear whether the factoring sequence is yet efficient or not when the
searching for the zero points starts at those time points $m$ far from the
zero points. The time development behavior of the long-range-interaction and
higher-order multiple-quantum spectral intensities may play an important
role to find efficiently the zero points when the searching for the zero
points starts at those points far from the zero points. The time development
behavior is dependent on the Hamiltonian $H(y,r,N)$ of the modular
exponential operation. The numerical simulation using the density operator $%
\rho _{y}(m)$ of Eq.(27) shows that the zero-quantum peak $I_{y}(p=0,m=1)$
is stronger than any other multiple-quantum peaks when the integer $%
y=2,4,...,$ and is much smaller than $N$. The simulation also shows that
some nonzero-order quantum peaks are also quite strong even for a large
integer $N$ and the smallest integer $y=2,$ although the zero-quantum peak
is still strongest. This implies that the Hamiltonian $H(y,r,N)$ with a
small integer $y=2,4,...$ and a large integer $N$\ still have a quite large
multiple-quantum coherence component $H_{1}$ in addition to the strongest
zero-quantum coherence operator $H_{0},$ so that the strong zero-quantum
peak falls off rapidly as the integer $m$ and the density operator $\rho
_{y}(m)$ approaches rapidly to its the steady state. However, the searching
for the zero points using the zero-quantum peak or other multiple-quantum
peaks is still locally efficient, that is, if the initial integer $m$ is
sufficiently near the zero points then the zero points can be found
efficiently even for a large number $N$ with the help of the time
development behavior of the zero-quantum peak or other multiple-quantum
peaks. The searching for the zero points based on the time development
behavior of the $p-$order quantum peak may become really inefficient when
the searching starts from those points $m$ far from the zero points. For an
integer $y\neq 2,4,...,$ or for a large integer $y$ the density operator $%
\rho _{y}(m)$ of Eq.(25) approaches quickly to its steady state as the
integer $m$. One of the reasons for it could be that the initial density
operator $\rho (0)$ of Eq.(21) is very special in the factoring sequence,
that is, $\rho _{S_{n}}(0)$ is the unity operator. It can be known from
Eq.(56) that if the initial density operator $\rho _{S_{n}}(0)$ is a $LOMSO$
operator then the multiple-quantum coherences created by the conditional
modular exponential operation $U_{I_{n_{i}}S_{n}}(y,r,N)^{m}$ belong to each
isolated subset $S(x_{l})\times S(x_{l})$ and the maximum number $(\thicksim 
\frac{1}{2}(r^{2}-r))$ of multiple-quantum transitions in the subset $%
S(x_{l})\times S(x_{l})$ usually is much less than the maximum number $\frac{%
1}{2}(4^{n}-2^{n})$ of multiple-quantum transitions of the subensemble $%
S_{n} $ with $n$ spins-1/2. In particular, the maximum number of
multiple-quantum transitions induced by the conditional modular exponential
operation on the initial density operator $\rho (0)$ of Eq.(21) is not more
than $(N-1)$ ($\frac{1}{2}2^{n}\leq N<2^{n}$) for any given integer $m$.
This can be seen from the density operator $\rho _{y}(m)$ of Eq.(27). This
number is greatly less than the maximum number $\frac{1}{2}(4^{n}-2^{n})$.
This could be one of the reasons why the density operator $\rho _{y}(m)$
(27) enters into its steady state rapidly as the integer $m$. In the future
it will be studied in detailed how the initial density operator $\rho
_{S_{n}}(0)$ is chosen properly to overcome the steady state problem.

A possible scheme to overcome the steady-state problem is described below
from the point of view of manipulating the Hamiltonian $H(y,r,N)$ of the
modular exponential operation. First one finds a unitary operator $G(y,r,N)$
so that the transformed Hamiltonian $\hat{H}(y,r,N)$ has a dominating
zero-quantum coherence component:

$\qquad \qquad \hat{H}(y,r,N)=G(y,r,N)^{+}H(y,r,N)G(y,r,N).$ \newline
Then this new Hamiltonian is acted on the initial density operator by
replacing the original Hamiltonian $H(y,r,N)$ of the modular exponential
operation, and the generated multiple-quantum spectra could be able to be
used to efficiently find the period $r$. The unitary operator $G(y,r,N)$
always exists, but it is a challenge how to find the exact unitary operation 
$G(y,r,N)$ that can be implemented in polynomial time. The Hamiltonian $%
H(y,r,N)$ always can be diagonalized unitarily. Assume that there is a
unitary operator $V(y,r,N)$ to diagonalize the unitary operator $U(y,r,N)$:

$\qquad V(y,r,N)^{+}U(y,r,N)V(y,r,N)=\Lambda (y,r,N).\qquad \qquad \qquad
\qquad \ \ (70)$ \newline
The unitary operator $V(y,r,N)$ can be constructed from the Fourier
transforms (12a) and (12b), but it contains the period $r$ in an explicit
form and this makes it difficult to construct its explicit quantum circuit.
But the unitary operator $V(y,r,N)$ could be built up approximately. The
Fourier transform (12a) over the period $r_{x_{l}}$ may be replaced with the
following approximated Fourier transform over the whole range of the integer 
$N$ [2, 3, 4, 6, 7]:

$\qquad |\Psi _{s}(x_{l})\rangle \thickapprox \frac{1}{\sqrt{N}}\stackrel{N-1%
}{\stackunder{k=0}{\sum }}\exp (i2\pi sk/N)|x_{l}y^{k}\func{mod}N\rangle .$
\qquad \qquad \qquad $\quad (71)$ \newline
If the period $r$ divides the integer $N$ then the Fourier transform (71) is
exact. But the period $r$ usually does not divide the integer $N$.
Therefore, the unitary operator $\hat{V}(y,N)$ built up with the Fourier
transform (71) diagonalizes approximately the unitary operator $U(y,r,N),$
that is, $V(y,N)\thickapprox \hat{V}(y,r,N)$. The unitary operator $\hat{V}%
(y,N)$ does not explicitly depend on the period $r$ and has a polynomial
quantum circuit since the Fourier transform (71) can be constructed
efficiently [2, 3, 4, 6, 7]. By using the unitary operator $\hat{V}(y,N)$ to
diagonalize approximately the Hamiltonian $H(y,r,N)$ the diagonal unitary
operator is obtained: $\hat{V}(y,N)^{+}U(y^{m},r,N)\hat{V}(y,N).$ Then one
chooses further a proper unitary operator $W(y,N),$ which Hamiltonian has a
dominating zero-quantum coherence component and a relative small
multiple-quantum component, to construct the desired unitary operator:

$\hat{U}(y,r,N)^{m}=W(y,N)^{+}\hat{V}(y,N)^{+}U(y^{m},r,N)\hat{V}%
(y,N)W(y,N).\qquad \ \ (72)$ \newline
Therefore, the unitary operator $G(y,r,N)$ that converts the Hamiltonian $%
H(y,r,N)$ into the desired Hamiltonian $\hat{H}(y,r,N)$ which has a
dominating zero-quantum coherence component may be approximated by the
unitary operator $G(y,N)$ which does not explicitly depend on the period $r$
and is given by

$\qquad G(y,r,N)\thickapprox G(y,N)=\hat{V}(y,N)W(y,N).$ \qquad $\qquad
\qquad \qquad \quad \ \ (73)$ \newline
Obviously, the unitary operator $\hat{U}(y,r,N)$ satisfies $\hat{U}%
(y,r,N)^{m}=E$ when $m=kr$ ($k=0,1,2,...,$)$,$ indicating that the unitary
operator $\hat{U}(y,r,N)$ has all the periods of the original unitary
operator $U(y,r,N)$. Generally the unitary operator $W(y,N)$ is chosen
suitably so that the unitary operator $\hat{U}(y,r,N)$ has not any other
periods except the own periods of the unitary operator $U(y,r,N)$. Now the
new unitary operation (72) places the original unitary operation $U(y,r,N)$
in the factoring sequence in previous sections. Then in the searching for
the zero points the initial density operator $\rho (0)$ of the spin ensemble
($I_{n_{i}}S_{n}$) is converted efficiently into the zero-quantum coherence
quickly at a short time, e.g., $m=1$, under the conditional unitary
operation $\hat{U}_{I_{n_{i}}S_{n}}(y,r,N)^{m}$ since the Hamiltonian $\hat{H%
}(y,r,N)$ has a dominating zero-quantum coherence component. Therefore, the
dominating zero-quantum coherence component of the Hamiltonian governs the
time evolution behavior of the spin ensemble at a short time and is
responsible for precisely distinguishing the zero points from other time
points in the time region near the zero points $\rho (kr)$ without an
exponential resource, while the relative small multiple-quantum coherence
component of the Hamiltonian will be responsible for the efficient searching
for the zero points starting from those time points far from the zero points.

According to the factoring sequence in previous sections one may find a zero
point $m=r^{\prime }$ by searching for the zero points of the density
operator $\rho _{y}(m)$, but the zero point $r^{\prime }$ could not be the
minimum period $r$ of the modular exponential function. Suppose the period $%
r $ is an even integer as before. Obviously, the ratio $r^{\prime }/r$ can
only take a half integer $(2k+1)/2$, an even integer $2k$, or an odd integer 
$(2k+1),$ $k=0,1,2,...$. If the ratio $r^{\prime }/r$ is a half integer,
i.e., $r^{\prime }/r=(2k+1)/2$ then the modular exponential function $%
f(y,r^{\prime },N)=y^{r^{\prime }}\func{mod}N=y^{r(2k+1)/2}\func{mod}%
N=y^{r/2}\func{mod}N.$ One can use directly the function $f(y,r^{\prime },N)$
to determine the non-trivial factor of the integer $N$ if the function $%
f(y,r^{\prime },N)\neq -1.$ The non-trivial factor takes either $\gcd
(f(y,r/2,N)-1,N)$ or $\gcd (f(y,r/2,N)+1,N)$ [1, 2, 3]$.$ Therefore, one
needs merely to find a half-integer zero point $r^{\prime }$ satisfying $%
r^{\prime }/r=(2k+1)/2$ to factor the integer $N$. Suppose that the
factoring sequence finds a zero point $r^{\prime }$. One uses the zero point 
$r^{\prime }$ to calculate the function $f(y,r^{\prime },N)$ which will take
about $O((\log _{2}N)^{3})$ steps [1, 2]$.$ If $f(y,r^{\prime },N)=1$ then $%
r^{\prime }/r=2k$ or $2k+1$, otherwise $r^{\prime }/r=(2k+1)/2.$ For the
case $r^{\prime }/r=2k$ or $2k+1$ one further calculates the function $%
f(y,r^{\prime }/2,N)$ by using the integer $r^{\prime }/2$, that is, $%
f(y,r^{\prime }/2,N)=y^{r^{\prime }/2}\func{mod}N.$ If now $f(y,r^{\prime
}/2,N)\neq \pm 1$ one can use the function $f(y,r^{\prime }/2,N)$ to find
further a non-trivial factor of the integer $N$; otherwise $f(y,r^{\prime
}/2,N)=1$ and $r^{\prime }/2=kr.$ For the case $r^{\prime }/2=kr$ one
calculates the function $f(y,r^{\prime }/4,N)$ again. If $f(y,r^{\prime
}/4,N)\neq \pm 1$ one will obtain a correct function $f(y,r^{\prime }/4,N)$
to factor the integer $N$, otherwise calculate further $f(y,r^{\prime }/8,N)$%
. Therefore, by $p=O(n)$ steps at most to calculate the function $%
f(y,r^{\prime }/2^{k},N),$ $k=0,1,2,...,p-1$ one can finally find a correct
function $f(y,r^{\prime }/2^{p-1},N)$ to factor the integer $N$. If the
period $r$ is not an even integer or $f(y,r^{\prime },N)=y^{r^{\prime }}%
\func{mod}N=-1$, meaning that one can not find a non-trivial factor of $N$
by the function $f(y,r^{\prime },N),$ one needs to choose another integer $y$
coprime to the integer $N$ [1, 2, 3] and then run the factoring sequence
above to find a zero point $r^{\prime }$ so as to obtain the correct $%
f(y,r^{\prime },N)$.\newline
\[
\]
$\qquad $\newline
{\large 6. Discussion}

In this paper a quantum factoring sequence based on the unitary dynamics of
quantum mechanics has been proposed to solve the prime factorization problem
on a spin ensemble without any quantum entanglement. It uses the NMR
multiple-quantum measurement techniques to output its quantum computational
results. The NMR\ quantum computer can perform the modular exponential
operation just like a classical digital computer, but its quantum
computational output is the inphase multiple-quantum spectrum of the spin
ensemble which may reduce merely in a polynomial form as the qubit number of
the spin ensemble. The computational complexity of the modular exponential
operation is the same on both the quantum computer and the classical one.
Quantum entanglement is not involved in the present ensemble quantum
computation of prime factorization because there is not any quantum
entanglement in the spin ensemble used to perform the prime factorization.
The time evolution process of the modular exponential operation on the
quantum computer obeys the unitary dynamics of quantum mechanics and hence
the computational output is governed by the Liouville-von Neumann equation
of quantum dynamics. This essential difference between the quantum computer
and the classical one could be the key point for the quantum computation
outperforming the classical one in the prime factorization on a spin
ensemble without any quantum entanglement. It has been shown that the prime
factorization based on the unitary dynamics of quantum mechanics on a spin
ensemble is locally efficient at least. Therefore, the quantum entanglement
could not be a unique resource to achieve speedup of quantum computation in
the prime factorization on a spin ensemble and quantum dynamics could play
an important role for the origin of power of quantum computation. The
steady-state problem is a harmful problem. It hampers the present factoring
sequence to find efficiently the period of the modular exponential function.
It is worth studying in detailed in the future how the steady-state problem
is dependent on the initial density operator of the factoring sequence and
how the steady-state problem may be overcome by manipulating the Hamiltonian
of the modular exponential operation.

There are a number of works [30, 31] to describe how to construct
efficiently the quantum circuit of the modular exponential unitary
transformation $U(y,r,N)$ in a quantum system with qubit number much more
than $(1+[\log _{2}N]),$ where a large number of extra auxiliary qubits are
used. The construction of the unitary operator $U(y,r,N)$ may be easier in a
quantum system with a larger Hilbert space, that is, with a larger number of
qubits. However, in practice it is still a challenge to construct
efficiently the quantum circuit of the unitary operator $U(y,r,N)$ in a spin
ensemble with $(1+[\log _{2}N])$ qubits at least. The implementation for the
Shor$^{\prime }$s factoring algorithm on a quantum system need consume a
number of qubits, but if the modular exponential operation could be
implemented efficiently on a spin ensemble with qubits as low as $(1+[\log
_{2}N])$ then this would simplify greatly the implementation of the prime
factorization.

Multiple-quantum coherences are generally measured indirectly through the
detection of single quantum coherence in NMR spectroscopy. The measurement
is more time-consuming than the direct detection of single quantum
coherence. However, the importance is that the measurement time for each
running of the factoring sequence based on the NMR multiple-quantum
spectroscopic method is almost independent of qubit number of a spin
ensemble. If digital resolution to record experimentally NMR
multiple-quantum signal needs to keep constant then the consuming time is
approximately linearly dependent on the qubit number since the spectral
width to cover over all $2n+1$ multiple-quantum spectral peaks is about $%
2n\omega _{S},$ approximately proportional to the qubit number $n$.
Therefore, the measurement is not a severe computational complexity problem
in the factoring sequence. An improved method to overcome the time-consuming
problem of multiple-quantum coherence indirect measurement might be using
one-dimensional multiple-quantum filtering experiments in NMR spectroscopy
[19]. The one-dimensional experiments should employ gradient magnetic field
[32] instead of the phase cycling to select the multiple-quantum coherence
with desired quantum order before detection and then convert it into single
quantum coherence to be detected directly. In the one-dimensional
experiments the antisymmetric property of the density operator $\rho _{y}(m)$
might be useful for the speedup of the searching for the zero points of the
density operator. However, there are some problems to be solved for the
one-dimensional experiments to be used in the NMR\ quantum computation, for
example, how to convert efficiently the desired order quantum coherences
into inphase single quantum coherence which can be detect efficiently.

Relaxation or decoherence effect in a spin ensemble is usually harmful for
any ensemble quantum computation, but it might be harmless for the searching
for the zero points in the factoring sequence on a spin ensemble. Since the
NMR multiple-quantum coherences usually decay in an exponential form and
irreversibly as the time development in a spin ensemble then the time
development behavior of the multiple-quantum spectral peaks such as the
zero-quantum peak used to search for the zero points may become more
distinct in the region near to the zero points so that the searching might
become more efficient. However, the decoherence effect may destroy the
efficient detection for multiple-quantum coherences and especially for those
higher-order quantum coherences due to the fact that a higher-order quantum
coherence usually has a shorter relaxation time and its NMR signal usually
decays much more rapidly than those lower order quantum coherences in a spin
ensemble. Therefore, relaxation effect is a compromised effect on the
present prime factorization on an NMR quantum computer. 
\[
\]
\newline
{\large References} \newline
1. P.W.Shor, Algorithms for quantum computation: Discrete logarithms and
factoring, Proc. 35th annual symposium on foundations of computer science,
IEEE computer society press, Los Alamitos, CA, pp. 124-134\newline
2. P.W.Shor, Polynomial-time algorithms for prime factorization and discrete
logarithms on a quantum computer, SIAM J.Comput. 26, 1484-1509 (1997)\newline
3. A.Ekert and R.Jozsa, Quantum computation and Shor$^{\prime }$s factoring
algorithm, Rev.Mod.Phys. 68, 733 (1996)\newline
4. R.Cleve, A.Ekert, C.Macchiavello, and M.Mosca, Quantum algorithms
revisited, Proc.R.Soc.Lond.A 454, 339 (1998)\newline
5. A.Kitaev, Quantum measurements and the Abelian stabiliser problem,
http://arxiv.org/abs/quant-ph/9511026 (1995)\newline
6. R.Jozsa, Quantum algorithm and the Fourier transform,

http://arxiv.org/abs/quant-ph/9707033 (1997) and Proc.R.Soc.Lond.A 454, 323
(1998)\newline
7. R.Jozsa, Quantum factoring, discrete logarithms and the hidden subgroup
problem, http://arxiv.org/abs/quant-ph/0012084\newline
8. L.M.K.Vandersypen, M.Steffen, G.Breyta, C.S.Yannoni, M.H.Sherwood, and
I.L.Chuang, Experimental realization of Shor$^{\prime }$s quantum factoring
algorithm using nuclear magnetic resonance, Nature 414, 883 (2001)\newline
9. A.Ekert and R.Jozsa, Quantum algorithms: entanglement-enhanced
information processing, Phil.Trans.Roy.Soc.Lond.A 356, 1769 (1998)\newline
10. R.Raz, Exponential separation of quantum and classical communication
complexity, Proc. 31st Annual ACM Symposium on Theory of Computing, 358
(1999)\newline
11. A.Ekert, Quantum cryptography based on Bell$^{\prime }$s theorem.
Phys.Rev.Lett. 67, 661 (1991)\newline
12. S.Lloyd, Quantum search without entanglement,

Phys.Rev. A61, 010301(R) (1999)\newline
13. D.A.Meyer, Sophisticated quantum search without entanglement.

Phys.Rev.Lett. 85, 2014 (2000)\newline
14. S.Parker and M.B.Plenio, Efficient factorization with a single pure
qubit and $\log _{2}N$ mixed qubits. Phys.Rev.Lett. 85, 3049 (2000)\newline
15. E.Knill and R.Laflamme, On the power of one bit of quantum information,
Phys.Rev.Lett. 81, 5672 (1998)\newline
16. S.Parker and M.B.Plenio, Entanglement simulations of Shor$^{\prime }$s
algorithm, http://arxiv.org/abs/quant-ph/0102136 (2001)\newline
17. S.Braunstein, C.M.Caves, R.Jozsa, N.Linden, S.Popescu, and R.Schack,
Separability of very noisy mixed states and implications for NMR quantum
computing, Phys.Rev.Lett. 83, 1054 (1999)\newline
18. A.Abragam, Principles of nuclear magnetism, (Oxford University Press,
London, 1961)\newline
19. R.R.Ernst, G.Bodenhausen, and A.Wokaun, Principles of Nuclear Magnetic
Resonance in One and Two Dimensions, (Oxford University Press, Oxford, 1987)%
\newline
20. R.Freeman, Spin Choreography, (Spektrum, Oxford, 1997)\newline
21. X.Miao, Universal construction of unitary transformation of quantum
computation with one- and two-body interactions, \newline
http://arXiv.gov/abs/quant-ph/0003068 (2000) \newline
22. X.Miao, Universal construction for the unsorted quantum search
algorithms, http://arXiv.org/abs/quant-ph/0101126 (2001)\newline
23. X.Miao, A polynomial-time solution to the parity problem on an NMR
quantum computer, http://arXiv.org/abs/quant-ph/0108116 (2001)\newline
24. X.Miao, Solving the quantum search problem in polynomial time on an NMR
quantum computer, http://arXiv.org/abs/quant-ph/0206102\newline
25. See for example, E.Jiang, K.Gao, and J.Wu, Linear algebra, (in Chinese),
(The People$^{\prime }s$ Education Press, 1978)\newline
26. Y.S.Yen and A.Pines, Multiple-quantum NMR in solids, J.Chem.Phys. 78,
3579 (1983) \newline
27. W.K.Rhim, A.Pines, and J.S.Waugh, Violation of the spin-temperature
hypothesis, Phys.Rev.Lett. 25, 218 (1970)\newline
28. Loo Keng Hua, An introduction to number theory, (in Chinese), (Science
Press, Beijing, 1957)\newline
29. X.Miao, Multiple-quantum operator algebra spaces and description for the
unitary time evolution of multilevel spin systems, Molec.Phys. 98, 625
(2000) \newline
30. D.Beckman, A.N.Chari, S.Devabhaktuni, and J.Preskill, Efficient networks
for quantum factoring, Phys.Rev. A 54, 1034 (1996) \newline
31. V.Vedral, A.Barenco, and A.Ekert, Quantum networks for elementary
arithmetic operation, Phys.Rev. A54, 147 (1996)\newline
32. J.Keeler, R.T.Clowes, A.L.Davis, and E.D.Laue, Methods in Enzymology,
Vol. 239, p. 145, (Academic Press, San Diego, 1994)\newline
\newline
\newline
\newline

\end{document}